\documentclass[twocolumn, twocolappendix]{aastex63}
\usepackage[italicdiff]{physics} 
\usepackage{tcolorbox} 

\usepackage[utf8]{inputenc}
\usepackage{amssymb,natbib,graphicx,color,ctable, amsmath,array,enumitem,kantlipsum}
\usepackage{hyperref}
\hypersetup{
    colorlinks=true,
    linkcolor=blue,
    citecolor=blue,
    filecolor=blue,
    urlcolor=blue
}

\usepackage[encapsulated]{CJK}
\usepackage{ucs}
\newcommand{\cntext}[1]{\begin{CJK}{UTF8}{bsmi}#1\end{CJK}}

\usepackage[flushleft]{threeparttable}
\usepackage{lineno}
\begin{document}

\defcitealias{HSvD21}{HSvD21}

\title{[CII] Emission in a Self-Regulated Interstellar Medium}
\author[0009-0004-2434-8682]{Alon Gurman}
\affiliation{School of Physics \& Astronomy, Tel Aviv University, Ramat Aviv 69978, Israel}
\author[0000-0002-9235-3529]{Chia-Yu Hu (\cntext{胡家瑜})}
\affiliation{Department of Astronomy, University of Florida, 211 Bryant Space Science Center, Gainesville, FL 32611, USA}
\affiliation{Max-Planck-Institut f\"{u}r Extraterrestrische Physik, Giessenbachstrasse 1, D-85748 Garching, Germany}
\author[0000-0001-5065-9530]{Amiel Sternberg}
\affiliation{School of Physics \& Astronomy, Tel Aviv University, Ramat Aviv 69978, Israel}
\affiliation{Center for Computational Astrophysics, Flatiron Institute, 162 5th Ave, New York, NY 10010, USA}
\affiliation{Max-Planck-Institut f\"{u}r Extraterrestrische Physik, Giessenbachstrasse 1, D-85748 Garching, Germany}
\author[0000-0001-7591-1907]{Ewine F. van Dishoeck}
\affiliation{Max-Planck-Institut f\"{u}r Extraterrestrische Physik, Giessenbachstrasse 1, D-85748 Garching, Germany}
\affiliation{Leiden Observatory, Leiden University, P.O. Box 9513, NL-2300 RA Leiden, the Netherlands}
\correspondingauthor{Alon Gurman}
\email{alongurman@gmail.com}

\begin{abstract}
The [CII] 157.74 $\mu$m fine structure transition is one of the brightest and most well-studied emission lines in the far-infrared (FIR), produced in the interstellar medium (ISM) of galaxies. We study its properties in sub-pc resolution hydrodynamical simulations for an ISM patch with gas surface density of $\Sigma_{\rm{g}}=10\;M_{\odot}\;\rm{pc}^{-2}$, coupled with time-dependent chemistry, far-ultraviolet (FUV) dust and gas shielding, star formation, photoionization and supernova (SN) feedback, and full line-radiative transfer. We find a [CII]-to-H$_2$ conversion factor that scales weakly with metallicity $X_{\rm{[CII]}}=6.31\times 10^{19} \;Z^{\prime\;0.17}\; \rm{cm}^{-2}\;(\rm{K}\;\rm{km}\;\rm{s}^{-1})^{-1}$, where $Z^{\prime}$ is the normalized metallicity relative to solar. {The majority of [CII] originates from atomic gas with hydrogen number density $n\sim 10~{\rm cm^{-3}}$.} The [CII] line intensity positively correlates with the star formation rate (SFR), with a normalization factor that scales linearly with metallicity. {We find that this is broadly consistent with $z\sim0$ observations.} As such, [CII] is a good SFR tracer even in metal-poor environments where molecular lines might be undetectable. Resolving the clumpy structure of the dense ($n=10-10^3\;\rm{cm}^{-3}$) interstellar medium (ISM) is important as it dominates [CII] 157.74 $\mu$m emission. We compare our full radiative transfer computation with the optically-thin limit and find that the [CII] line becomes marginally optically thick only at super-solar metallicity for our assumed gas surface density. 
\end{abstract}




\section{Introduction}
Singly ionized carbon (C$^+$) is an important component of the interstellar medium (ISM). Given its low ionization potential of 11.26 eV, it is abundant in the warm- and cold neutral ISM (WNM, CNM), as well as part of the molecular ISM, as the carbon is ionized by far-ultraviolet (FUV) radiation longward of the Lyman limit which escapes regions of ionized gas surrounding hot OB stars. It also resides in HII regions alongside higher ionization stages of carbon. C$^+$ is an important coolant in the cold ISM, via fine structure line emission at 157.74 $\mu$m \citep[hereafter {[}CII{]} emission;][]{Field1969, Draine1978, Crawford1985,Wolfire2003,Bialy2019}. [CII] is the strongest spectral line in the far-infrared in most galaxies. It radiates away energy injected into the neutral atomic components of the ISM, mostly by FUV dust photo-electric heating. The upper level of the transition is excited by collisions, mostly with neutral and molecular hydrogen in the neutral ISM, and with electrons in HII regions. At sufficient densities, [CII] emission allows gas to cool down to temperatures of $\sim$10$^2$ K, owing to its low excitation threshold of 91.21 K \citep{Dalgarno1972}. [CII] emission is thus important for the thermodynamics of star formation, as gas cools and collapses under its self-gravity. [CII] emission is therefore ubiquitous in observations of stellar nurseries in our galaxy, as well as tracing star formation in extragalactic sources. 

In recent years, observations of [CII] emission have been pushed to higher redshifts, reaching as high as $z\sim7$ {\citep{Maiolino2015,Knudsen2016,Carniani2018,Hashimoto2019,Matthee2019}}, mainly due to the operation of large millimeter wave interferometers such as the Atacama Large Millimeter/submillimeter Array (ALMA) and the Northern Extended Millimeter Array (NOEMA). Observations have established a correlation between star formation rate (SFR) and [CII] line luminosity. At redshifts $z\approx 0$ the Infrared Space Observatory \citep{Delooze2011} and the Herschel Space Observatory \citep{DeLooze2014, HerreraCamus2015} first established the [CII]--SFR relation on both galactic scales as well as on a spatially resolved level. On galactic scales, this relationship seems to hold for higher redshift systems as well, as revealed by ALMA up to $z\sim7$ \citep{LeFevre2020, Schaerer2020,Bouwens2022,Harikane2020}. {\cite{DeLooze2014} found that the scatter in the [CII]--SFR relation is correlated with metallicity. The lowest metallicity galaxies in the sample by \cite{HerreraCamus2015}, dominated by galaxies with a metallicity (relative to solar) of $\sim0.5$, showed an offset from their observed [CII]--SFR relation. These results point to the possibility of a metallicity dependent [CII]--SFR relation. Even so, the scatter in the high-$z$ observations is larger and a metallicity trend in the data is unclear.}

Most of the cold gas mass in the ISM is in the form of H$_2$, but due to its lack of permanent dipole moment, it does not radiate efficiently at the low ($T\lesssim 10^2$ K) temperatures of the cold ISM. [CII], alongside other emission lines (especially CO rotational emission), is therefore one of the main tracers of cold gas in galaxies. In cold, dense gas that is shielded from FUV radiation, C$^+$ can recombine efficiently to form C, which in turn forms CO in even more shielded regions of gas clouds \citep{Tielens1985, Sternberg1995, Wolfire2022}. At the same time, dense HII regions, which have a significant fraction of their carbon in C$^+$, also show strong [CII] emission \citep{Heiles1994}. This adds a layer of complexity to tracing cold gas and star formation using the [CII] line, as the origin of the emission needs to be understood to properly interpret observational data. 

Traditionally, CO line emission is used to trace cold molecular gas. Specifically, the intensity of the CO $J=1-0$ line at 2.6 mm can be measured, and converted into H$_2$ column density using the CO-to-H$_2$ conversion factor defined by

\begin{equation}
    X_{\rm{CO}}\equiv \frac{N_{\rm{H}_2}}{W_{\rm{CO}}},
\end{equation}
where $N_{\rm{H}_2}$ is the H$_2$ column density, and $W_{\rm{CO}}$ is the velocity integrated brightness temperature of the CO $J=1-0$ line, measured in units of K km s$^{-1}$. One can similarly define a [CII]-to-H$_2$ conversion factor

\begin{equation}
\label{eq: XCII}
    X_{\rm{[CII]}}\equiv \frac{N_{\rm{H}_2}}{W_{\rm{[CII]}}},
\end{equation}
where $W_{\rm{[CII]}}$ is the velocity-integrated brightness temperature of the [CII] line. {Analogously, one can define a conversion factor between the line luminosity and molecular mass}
\begin{equation}
    \alpha_{\rm{CO}}=\frac{M_{\rm{H}_2}}{L^{\prime}_{\rm{CO}}},
\end{equation}
{where $M_{\rm{H}_2}$ is the H$_2$ mass and $L^{\prime}_{\rm{CO}}$ is the area-integrated value of $W_{\rm{CO}}$, in units of K km s$^{-1}$ pc$^2$. The equivalent factor $\alpha_{\rm{[CII]}}$ is defined similarly, with $L^{\prime}_{\rm{CO}}$ replaced by $L_{\rm{[CII]}}$, i.e., the luminosity.} The potential and applicability of such a conversion factor ties into the question of the CO-dark H$_2$, i.e., molecular gas not traced by CO emission. CO-dark molecular gas is a result of the HI-to-H$_2$ transition taking place at lower column densities than the C$^+$-C-CO transition, and the fraction of H$_2$ in this intermediate layer varies with gas properties, most importantly with gas metallicity, as the CO core becomes smaller and might even be absent entirely \citep{Wolfire2010b, Nordon2016, Hu2022a}. This question has been investigated observationally \citep{Langer2010,Langer2014,Pineda2013}, as well as using 1D photodissociation region (PDR) models \citep{Tielens1985, Pak1998, Hollenbach1999, Wolfire2022}. \cite{Madden2020} used PDR modelling to analyze observations from the Dwarf Galaxy Survey \citep{Madden2013}, {and found a relation between H$_2$ mass and [CII] line luminosity of the form $M_{\rm{H}_2}=10^{2.12}\times (L_{\rm{[CII]}}/L_{\odot})^{0.97} M_{\odot}$, i.e., an approximately metallicity-independent $\alpha_{\rm{[CII]}}=132\;M_{\odot}\;L_{\odot}^{-1}$. \cite{Zanella2018} complement their $z\sim 1-2$ [CII] observations with data from the literature spanning redshifts $z\sim0-5$ and including both main-sequence and starburst galaxies, and find  $\alpha_{\rm{[CII]}}=31\;M_{\odot}\;L_{\odot}^{-1}$ with a standard deviation of 0.3 dex. Their estimation of H$_2$ mass was done either by using the models of \cite{Sargent2014} and \cite{Scoville2017} where CO observations were unavailable, or using $X_{\rm{CO}}$ of the form}
\begin{equation}
    X_{\rm{CO}}=X_{\rm{CO,MW}}\times(Z^{\prime}/Z^{\prime}_{\odot})^{-1.5},
\end{equation}
{where $X_{\rm{CO,MW}}$ is the CO-to-H$_2$ conversion factor observed in the Milky Way. This relation between $X_{\rm{CO}}$ and metallicity is notably flatter than the $Z^{\prime-3.39}$ dependence found by \cite{Madden2020}, and steeper than the $Z^{\prime-0.71}$ found by \cite{Hu2022a}.}

Indeed, theoretical models of 1-dimensional photodissociation regions (PDRs) can be used to quantitatively model the connection between gas properties and [CII] emission. In recent years, however, the study of the ISM emission in 3D hydrodynamical numerical simulations has been made possible thanks to the advent of large supercomputers. Simulations can now track gas dynamics, thermal state, chemistry (albeit usually in post-processing), and the different star formation feedback mechanisms, to give a more realistic context to the theoretical study of ISM properties. Hydrodynamical simulations have attempted to capture different spatial scales, from single cloud, through galactic, to cosmological scales {\citep[e.g.,][]{Glover2011,Vallini2013,Walch2015,Olsen2015,Hu2016,Hu2017,Kim2017,Lupi2017,Katz2017,Katz2019,Pallottini2019,Pallottini2022} }.

To capture the cold gas component of the ISM in a realistic environment, however, one must spatially resolve the scale of dense molecular cores, as well as have the spatial and temporal extent that captures the larger WNM gas reservoir and the effects of stellar feedback due to photoionization and supernovae (SNe). The emerging life cycle of the ISM includes the cooling and gravitational collapse of gas clouds to form stars, followed by heating, ionization, dispersal, and injection of turbulence into the gas by massive stars and SNe. Simulations of [CII] emission vary in scale, hydrodynamical setup, implementation of sub-grid physical models, and treatment of radiative transfer. 

{\cite{Hu2017} carried out a simulation of an isolated dwarf galaxy at high spatial resolution, computing unresolved [CII] emission under the optically thin assumption. \cite{Lupi2020} simulated a dwarf galaxy using the zoom-in technique, where the boundary conditions of gas flow are determined by a larger, lower resolution cosmological simulation, and again calculated the unresolved, optically thin, [CII] emission. \cite{Bisbas2022} simulated the merger of two dwarf galaxies and post-processed their simulation results with a full radiative transfer computation. They found that [CII] emission originates mostly from the WNM, and reproduced the [CII]--SFR relation observed in nearby dwarf galaxies.} 

{Other groups used various treatments of chemistry, (F)UV radiative transfer, ISM sub-grid structure, and [CII] line emission in a cosmological context, for either a large sample of galaxies or a sub-sample of galaxies using the zoom-in technique. \cite{Liang2023} modelled each particle in the FIRE-2 suite of zoom-in cosmological simulations using tabulated CLOUDY photoionization models \citep{Ferland2017} and obtained [CII] intensities using the optically thin assumption. \cite{Olsen2015,Olsen2017,Leung2020} used an approximation for the local stellar FUV radiation with a more elaborate set of CLOUDY models, for both ionized and neutral gas. \cite{Lupi2020b,Katz2019,Katz2022,Pallottini2019,Pallottini2022} treated (F)UV radiative transfer using M1 closure based schemes \citep{Aubert2008}.
\cite{Vallini2013,Vallini2015} used a Monte-Carlo radiative transfer method to treat UV radiation, as well as a sub-grid model for the unresolved ISM density structure. Their sub-grid model accounts for emission from the CNM, WNM, molecular clouds, and the PDRs surrounding them. They found that $\lesssim 50\%$ of [CII] emission originates in WNM and CNM, and this fraction to be anti-correlated with SFR. \cite{Pallottini2019,Pallottini2022} also applied a sub-grid model for unresolved ISM structure, and found that [CII] emission originates mostly from gas with densities above $1-10$ cm$^{-3}$, depending on the adopted model and metallicity. \cite{Vizgan2022} used a sample of $\sim1000$ galaxies in a cosmological simulation, coupled with sub-grid models for both [CII] luminosity and H$_2$ fraction and found a sublinear $M_{\rm{H}_2}-L_{\rm{[CII]}}$ relation with median value of $\alpha_{\rm{[CII]}}=18\;M_{\odot}\;L_{\odot}^{-1}$, notably lower than the value of $31\;M_{\odot}\;L_{\odot}^{-1}$ found by \cite{Zanella2018}. They also concluded that $\gtrsim 50\%$ of the [CII] emission originates from molecular gas, and that this fraction increases with galaxy stellar mass.} 

\cite{Franeck2018, Ebagezio2022} used an analogue of the zoom-in technique but zooming in on a single cloud within a simulation of a patch of a galactic disk. Their simulations have very high spatial resolution, include non-equilibrium chemistry and cooling, and they post-process their results using photoionization modelling in order to account for stellar photoionization feedback. They also perform full radiative transfer calculations in order to obtain synthetic [CII] emission maps. Their high resolution means that they simulate a relatively small volume, for a simulation time shorter than the typical lifespan of a gas cloud, and for solar metallicity. They quantified $X_{\rm{[CII]}}$ for each collapsing molecular cloud, and found values ranging from $\sim0.5$ to $\sim12\times$10$^{20}$ cm$^{-2}$ (K km s$^{-1}$)$^{-1}$. The range in values of $X_{\rm{[CII]}}$ is due to both time variation and variation between clouds and simulation setup. 


While most simulations investigating the [CII]--SFR relation succeed in reproducing observations, albeit to different levels of success, they usually use either a constant metallicity or couple the metallicity evolution to the SFR in the galaxy. \cite{Lagache2018} used a semi-analytic approach to model the [CII]--SFR relation, and found a wide correlation between [CII] luminosity and metallicity, with a very large scatter of $~0.8$ dex. \cite{Vallini2015} varied the metallicity in a single snapshot of a zoom-in cosmological simulation {where they accounted for contribution from PDRs using a sub-grid model,} and found a positive trend in the [CII]--SFR ratio with metallicity. \cite{Olsen2015} found a weaker trend in the [CII]--SFR ratio with metallicity, but did not rule it out, as their results spanned a range of just over 0.5 dex in carbon abundance. {In the high-resolution zoom simulations of \cite{Lupi2020}, the authors reproduced the positive correlation in the [CII]--SFR ratio found by \cite{Vallini2015}. \cite{Liang2023} found that the metallicity is the main driver of the variation in the [CII]--SFR relation for $z\geq4$ galaxies, but also claimed that the metallicity effect is cancelled out at lower redshifts by a decrease in the fraction of carbon mass in the form of C$^+$ due to conversion to C and CO.}

\citet{HSvD21} (hereafter \citetalias{HSvD21}) performed high-resolution simulations of an SN driven, self-regulated ISM, with a mass resolution of 1 $M_{\odot}$, spatial resolution of $\sim$0.2 pc, and a timespan of 500 Myr. A wide range of metallicities was considered from $Z^\prime=0.1$ to 3,  where $Z^\prime \equiv Z/Z_{\odot}$ is the gas metallicity relative to solar. Their simulations cover the spatial range of dense molecular cores, and for a long enough simulation time such that clouds are allowed to form and disperse over several cycles. To investigate [CII] emission, radiative transfer calculations must be carried out. This involves calculating the excitation state of each gas particle, and solving the radiative transfer equation accounting for both emission and absorption, to create synthetic emission maps for the simulation snapshots.

This paper is organized as follows. In Section \ref{sec: numerics}, we briefly describe our numerical setup for both the \citetalias{HSvD21} hydrodynamical simulations, and radiative transfer calculation. In Section \ref{sec: carbon chem} we discuss the chemical and excitation physics relevant to [CII] emission. In Section \ref{sec: overview}, we present an overview of our results, in the form of spatially resolved column density and emission maps, spatially and temporally averaged quantities (and their metallicity dependence), and the chemical and excitation state of our gas. In Section \ref{sec: time var}, we discuss the time variation of SFR, H$_2$ column density, and [CII] emission, and present our $L_{[CII]}$--SFR relation. In Section \ref{sec: optical depth}, we discuss the effect of optical depth on our results. In Section \ref{sec: summary}, we summarize our work.

\section{Numerical Methods}
\label{sec: numerics}
\subsection{Simulations}
The hydrodynamical simulations we analyze in this paper are from \citetalias{HSvD21},
which we summarize as follows.
The simulation setup is the so-called ``stratified box''.
It has a size of 1~kpc in both $x$- and $y$-axes with periodic boundary conditions
and 10~kpc in the $z$-axis with outflow boundary conditions.
The center of the box is the origin and the mid-plane of the disk is located at $z = 0$.
Gas is vertically distributed such that the thermal pressure balances 
the self-gravity of gas and the additional gravity of a stellar disk and the dark matter halo.
The simulation is conducted using the public version of {\sc Gizmo} \citep{Hopkins2015}
which uses a meshless Godunov method \citep{Gaburov2011} 
and is built on the TreeSPH code {\sc Gadget-3} \citep{Springel2005}.
Time-dependent cooling and H$_2$ chemistry are included
based on \citet{Glover2007} and \citet{Glover2012},
with a {\sc HealPix} \citep{Gorski2011}-type treatment for radiation shielding.
Star formation is included using the standard stochastic approach 
that depends on the local free-fall time,
with a star formation efficiency of 50\%.
The stellar mass of a massive star is drawn from an initial stellar mass function of \citet{Kroupa2002}
that determines its stellar lifetime and luminosity of ionizing radiation. 
Stellar feedback includes supernovae (SNe) and photoionization following the method of \citet{Hu2017}.
The FUV radiation field and cosmic-ray ionization rate 
both scale linearly with the total star formation rate
and are assumed to be spatially uniform.
The metallicity is assumed to be constant in each simulation.

A broad range of metallicity is explored 
with $Z^\prime = 3, 1, 0.3, \text{and } 0.1$,
each simulation is run for 500~Myr.
The results are post-processed to calculate the chemical species
of C$^+$, C, and CO in all gas cells
using {\sc AstroChemistry.jl}\footnote{Code is publicly available at \url{https://github.com/huchiayu/AstroChemistry.jl}} {\citep{Hu2021}},
taking the time-dependent abundances of H$_2$ and H$^+$ as known parameters.
Shielding against the FUV radiation is calculated with a {\sc HealPIX}-based method
that accounts for shielding by dust, H$_2$, and CO.

\subsection{Radiative Transfer on an Adaptive Mesh}

Following \citet{Hu2022a},
we generate synthetic [CII] emission maps using the radiative transfer code {\sc Radmc-3D} \citep{Dullemond2012}.
For each snapshot, 
we generate $512 \times 512$ spectra of [CII] covering the entire 1 kpc$^2$ simulation area
such that each pixel has a spatial resolution (pixel size) of $1~{\rm kpc} / 512 \approx 2$~pc.
Each spectrum has a spectral coverage of $\pm$20 ${\rm km~s^{-1}}$ sampled at 100 equally spaced wavelengths.
We do so for 41 snapshots from $200$ to $600$~Myr with a time interval of $10$~Myr.

Our spatial resolution is fully adaptive and Jeans mass-resolved down to $\sim 0.2$~pc.
Therefore,
it is computationally impractical to adopt a uniform mesh to cover our entire simulation domain.
We use {\sc ParticleGridMapper.jl}\footnote{Code is publicly available at \url{https://github.com/huchiayu/ParticleGridMapper.jl}}
to interpolate particle data onto an adaptive mesh with 13 refinement levels
and a minimal cell size of 0.12 pc.
As {\sc Radmc-3D} has the capability of solving the radiative transfer equations
on an adaptive mesh using the ``recursive sub-pixeling'' technique,
it guarantees that all [CII] emission can be properly captured even with a coarse pixel.

We adopt the non-LTE mode of radiative transfer in {\sc RadMC-3D}.
To account for radiation trapping,
we use the large-velocity gradient approximation implemented by \citet{Shetty2011} 
and calculate the level populations at each cell. 
The collision partners of C$^+$ are H, H$_2$ with an ortho-to-para ratio of 3, and $e^{-}$.

\label{subsec: RADMC}

\section{Carbon Chemistry}
\label{sec: carbon chem}
\subsection{C$^+$/C/CO Structure}

Most carbon in the neutral ISM is in the form of C$^+$. Lyman continuum (LyC) photons are absorbed in the vicinity of OB stars, but the ionization energy of carbon is 11.26 eV, lower than that of hydrogen. Dense, shielded regions of the cold ISM contain a significant mass of C and CO, forming a layer-like structure, with a transition from C$^+$, to C, to CO, going from the outer regions of a cloud inwards. The C$^+$ abundance relative to the total carbon abundance is close to unity in the WNM and unshielded CNM. The C$^+$/C ratio is set by a balance between FUV photoionization and C$^+$ recombination, either directly with electrons or in the form of grain-assisted recombination. Given a large enough column density, the photoionization rate drops via the processes of dust-shielding and H$_2$ shielding. This allows the formation of a layer of gas where carbon is in the form of C. In the inner regions of the cloud, where C ionizing radiation has been fully absorbed, the C/CO ratio is set by CO formation (via different intermediary molecules such as CH, OH, and O$_2$), and destruction via FUV photodissociation by photons with energy greater than 11.09 eV and the cosmic-ray (CR) destruction channel. With sufficient shielding from dust, H$_2$, and CO, carbon gas at the dense core of the cloud is fully converted into CO. In 1-dimensional PDR models, the physical parameters setting the C$^+$/C/CO transition structure are the Lyman-Werner band (11.2-13.6 eV) flux, gas density (or density profile), primary cosmic ray flux, and gas metallicity. 

The treatment of chemistry in the \citetalias{HSvD21} simulations is a hybrid time-dependent/steady-state approach, where a limited number of hydrogen chemical reactions are incorporated in an on-the-fly treatment, while the abundances of additional species are computed in post-processing and assuming steady-state. The chemical reaction rate coefficients are taken from the UMIST database \citep{McElroy2013}, while the photo-reactions are calculated using a HEALPIX 12-ray approximation to compute an effective shielding column for each gas particle. The photodissociation rate of CO (and similarly C and H$_2$) is then computed by the expression
\begin{equation}
    \Gamma_{\rm{CO}}=I_{\rm{UV}}\Gamma_{0,\rm{CO}}f_{\rm{dust,CO}}f_{\rm{gas,CO}},
\end{equation}
where $I_{\rm{UV}}$ is the FUV flux normalized to the Draine field, $\Gamma_{0,\rm{CO}}=2.43\times 10^{-10}$ s${^{-1}}$ is the unattenuated dissociation rate. $f_{\rm{dust,CO}}$ is the dust shielding factor, and $f_{\rm{gas,CO}}$ is the shielding factor from H$_2$ and CO self-shielding, both of which are computed using the effective H$_2$ and CO column densities from the HEALPIX computation. This setup has been shown to be consistent with 1D PDR models by \citetalias{HSvD21}.

\subsection{C$^+$ Excitation} 

\begin{figure}	
	\centering
	\includegraphics[width=1\columnwidth]{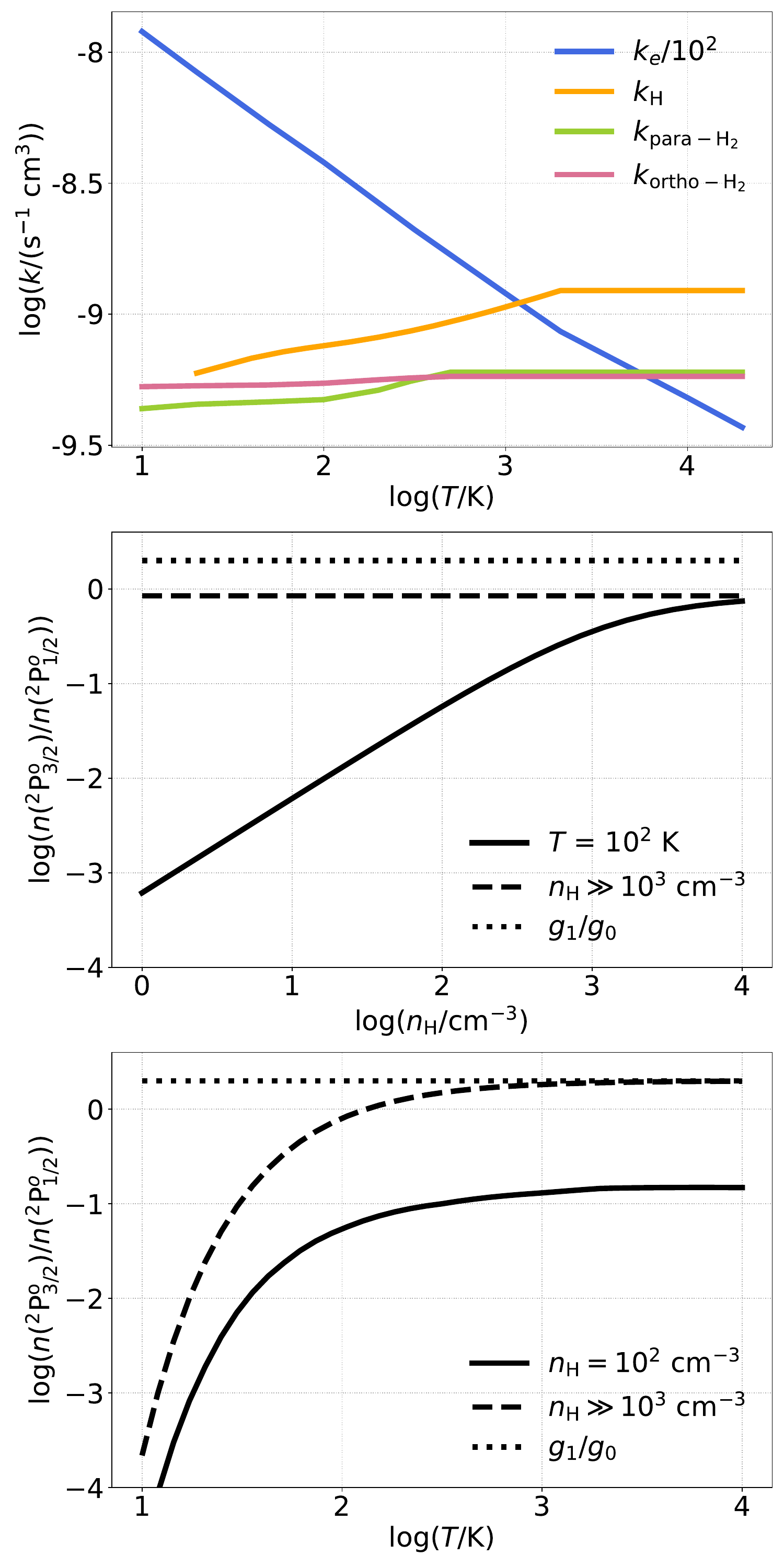} 
	\caption{
Top panel - collisional rate coefficients from the LAMDA database for electrons, H, ortho- and para-H$_2$. Middle panel - the relative abundance of the excited C$^+$ level $^2 \rm{P}_{3/2} ^{\rm{o}}$ as a function of density for a fixed temperature of $T=100$ K, assuming abundances of $x_{\rm{H}}=0.5$, $x_{\rm{e}}=10^{-4}$, $x_{\rm{H}_2}=0.25$, the ratio for a thermalized population $g_1/g_0 \times e^{-T_{10}/T}$, and the high temperature limit $g_1/g_0$. Bottom panel - as for the middle panel but showing the temperature dependence for a fixed H nucleus density.
		}
		\label{fig: collisional excitation}
\end{figure}

[CII] emission arises from the fine structure transition between the $ ^2 \rm{P}^{\rm{o}} _{3/2}$ and $^2 \rm{P}^{\rm{o}} _{1/2}$ of the ground state 1$s^2 2s^2 2p\; ^2 \rm{P}^{\rm{o}}$ of C$^+$. Excitation is due to collisions with electrons, H, and ortho- and para-H$_2$. The collisional rate coefficients used in this work are taken from the Leiden Atomic and Molecular Database \citep[LAMDA;][]{Schoier2005,Wilson2002,Barinovs2005,Lique2015}, and are presented in the top panel of Figure \ref{fig: collisional excitation}. The critical density for the transition in neutral gas is of order $10^3$ cm$^{-3}$, depending on gas temperature and chemical composition, and ~50 cm$^{-3}$ in ionized gas \citep{Galli2013}. The difference between the two values for neutral and ionized gas is due to the rate coefficient for collisions with electrons at 10$^4$ K being almost two orders of magnitude higher than that of H or H$_2$ at 10$^2$ K. Collisions with electrons are less important in neutral gas where the electron abundance is typically $\lesssim10^{-4}$, and collisions with H and H$_2$ become important instead.

The middle panel of Figure \ref{fig: collisional excitation} shows the density dependence of the excitation state of C$^+$ for a fixed temperature of 100 K, assuming abundances of $x_{\rm{H}}=0.5$, $x_{\rm{e}}=10^{-4}$, $x_{\rm{H}_2}=0.25$, and an orth-to-para H$_2$ ratio of 3, representing typical conditions in the cold ISM. Also plotted is the asymptotic value for $n\gg 10^3$ cm$^{-3}$, given by the Boltzmann factor $g_1/g_0\times \exp{(-T/T_{10})}$, and the high-temperature limit $g_1/g_0$, where $g_1$ and $g_0$ are the statistical weights for the upper and lower excitation states, respectively, and $T_{10}$ is the temperature corresponding the transition energy. The density dependence of the excitation state begins flattening at $n\sim 10^3$ cm$^{-3}$, corresponding to the critical density of the transition. The bottom panel of Figure \ref{fig: collisional excitation} shows the temperature dependence of the excitation state of C$^+$ for the same chemical abundances, for a density of $n=10^2$ cm$^{-3}$ and $n\gg 10^{3}$ cm$^{-3}$, as well as the high-temperature limit. The temperature dependence is weak when $T\gtrsim 100$ K, corresponding to the temperature of the transition at 91.21 K.

\begin{figure}
	\centering
    \includegraphics[trim=0.5cm 0cm 0cm 0cm,clip, width=1\linewidth]{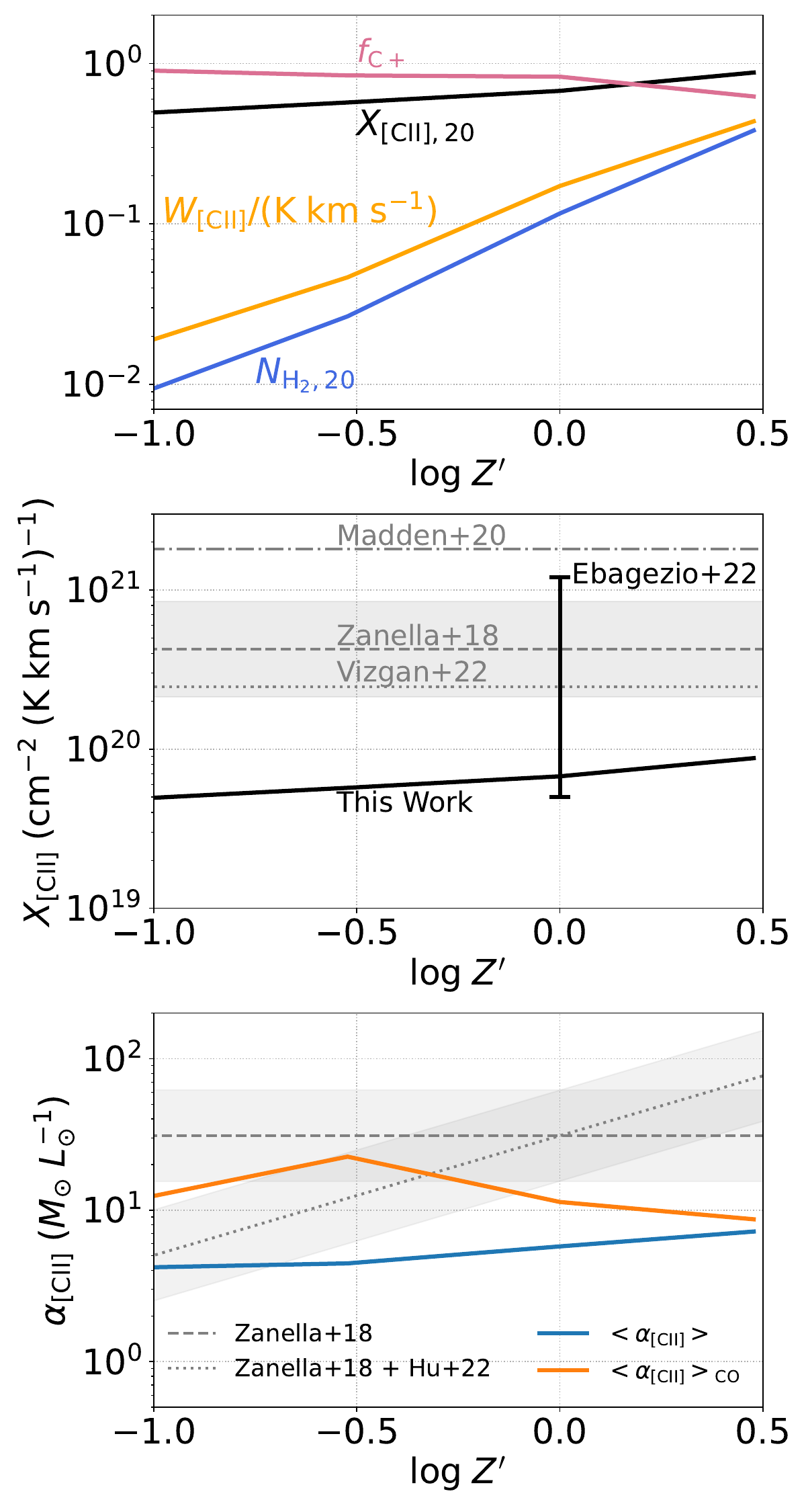} 
	\caption{
{Top panel: time average of $W_{\rm{[CII]}}$ (K km s$^{-1}$), $X_{\rm{[CII]},20}\equiv X_{\rm{[CII]}}/(10^{20}\;\rm{cm^{-2}\;(K\;km\;s}^{-1})^{-1})$, $N_{\rm{H}_2,20}\equiv N_{\rm{H}_2}/(10^{20}\;\rm{cm}^{-2})$, and $f_{\rm{C}^+}\equiv M_{\rm{C}^+} / {M_{\rm{C,tot}}(n>10\;\rm{cm}^{-3})}$. Middle panel: comparison between our computed $X_{\rm{[CII]}}$ (black line) and the range of values computed by \cite{Ebagezio2022} (vertical black line), the $\alpha_{\rm{[CII]}}$ value computed in \cite{Vizgan2022} (dotted grey line), the value derived in \cite{Madden2020} (dash-dotted line grey), and the observed value from \cite{Zanella2018} (dashed grey line) with their $1\sigma$ scatter in shaded grey, all in terms of $X_{[\rm{CII}]}$. Bottom panel: metallicity dependence of time averaged $\alpha_{\rm{[CII]}}$, as well as the $W_{\rm{CO}}$ weighted time average. Also shown is the constant found in \cite{Zanella2018}, as well as their value with a correction factor to account for the overestimation of H$_2$ as discussed in \cite{Hu2022a}}.
		}
		\label{fig: Z dependence}
\end{figure}

\section{Overview of Simulation Results}
\label{sec: overview}

\begin{figure*}
	\centering
	\includegraphics[trim=1cm 0cm 1cm 0cm,clip, width=1\linewidth]{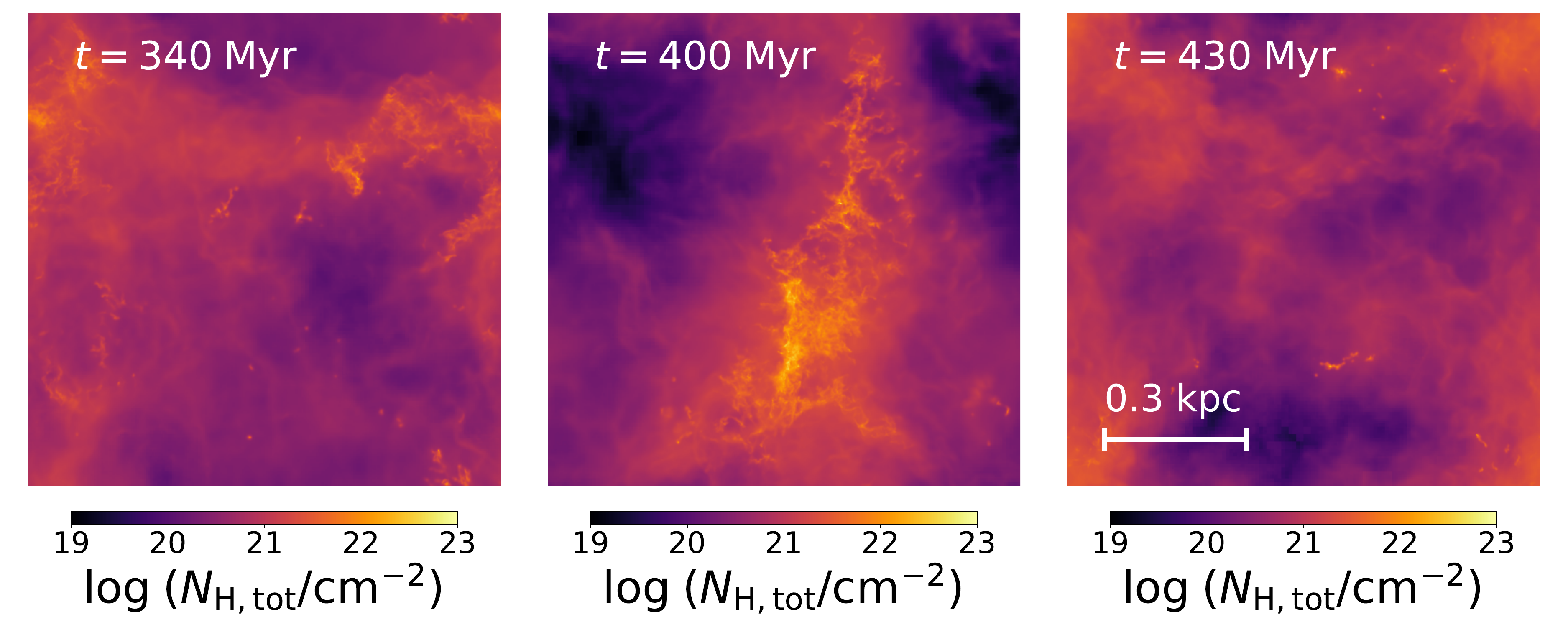} 
	\caption{
Maps of total column density for the $Z'=1$ run of the \citetalias{HSvD21} simulations, taken at $t=340$, 400, and 430 Myr, showing the effect of time evolution.
		}
		\label{fig: Z=1 snapshots}
\end{figure*}

\begin{figure*}
	\centering
    \centerline{\includegraphics[width=1\linewidth]{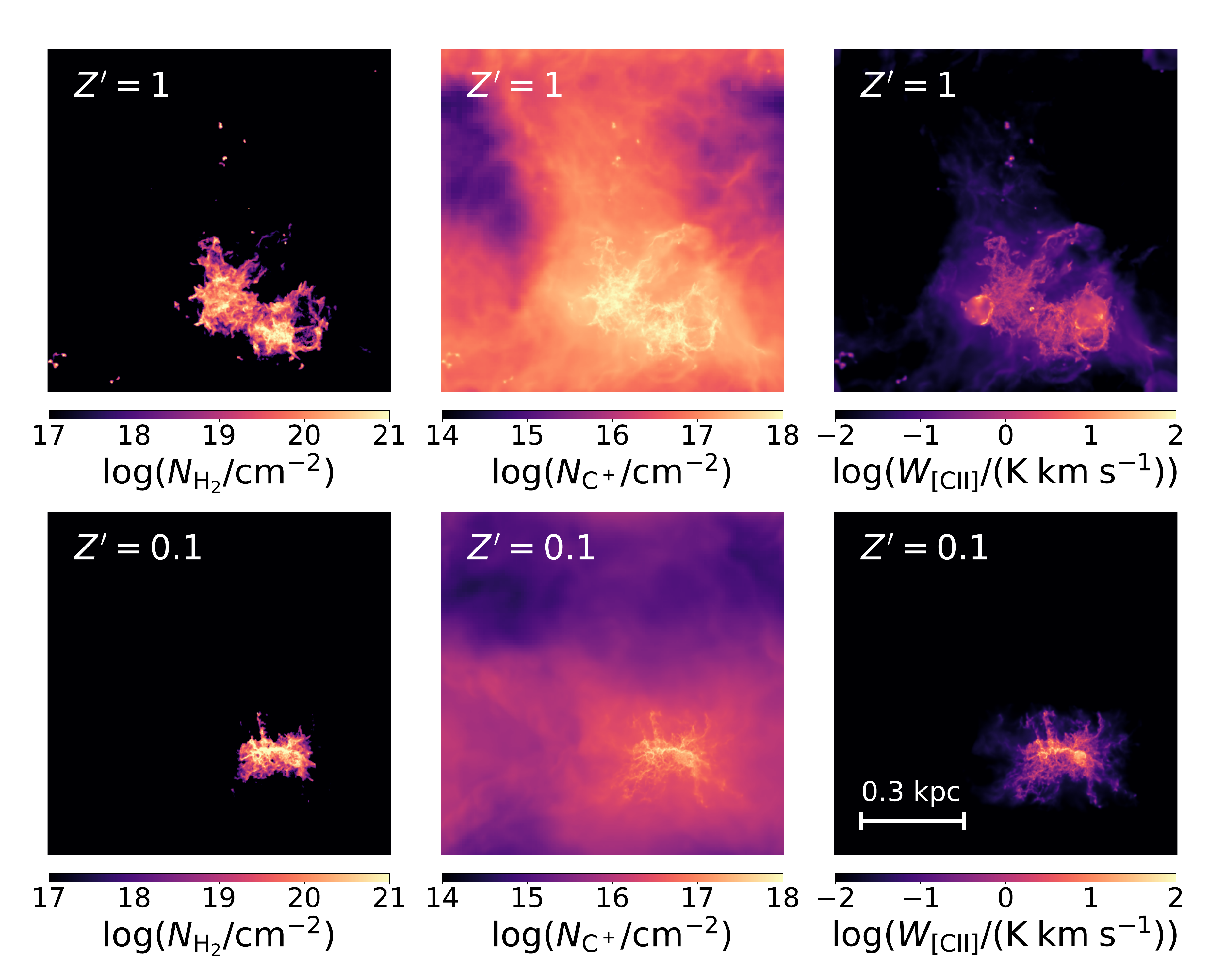}}
	\caption{
H$_2$ and C$^+$ column density maps, and [CII] emission maps for $Z'=1$ (top row) and 0.1 (bottom row), for $t=520$ and 230 Myr, respectively. While the C$^+$ distribution is extended, [CII] emission and H$_2$ both trace dense gas.
		}
		\label{fig: N and W maps}
\end{figure*}

\subsection{Metallicity Dependence}
\label{sec: metallicity dependence overview}
{The top panel of} Figure \ref{fig: Z dependence} shows the metallicity dependence of the time averaged H$_2$ column, $[\rm{CII}]$ intensity $W_{\rm{[CII]}}$, the conversion factor $X_{\rm{[CII]}}$ (see Equation \ref{eq: XCII}), and the mass-weighted mean C$^{+}$ abundance (normalized to the total carbon abundance) in $n>10\;\rm{cm}^{-3}$ gas, denoted as $f_{\rm{C}^+}$. The almost linear rise of $W_{\rm{[CII]}}$ can be understood by the fact that the emission is approximately optically thin, C$^+$ mass scales approximately linearly with metallicity, {and the shape of the gas density probability density function (PDF) is metallicity independent (see \citetalias{HSvD21})}. As is described in further detail in Section \ref{subsec: chemical properties}, [CII] emission in our calculations mostly originates {from moderately dense gas with $n\sim10 ~\rm{cm}^{-3}$, and safely below the conversion points from C$^+$ to C for all metallicities.} As the metallicity is increased, the abundance of C and CO increases due to an increase in dust shielding, and $f_{\rm{C}^+}$ decreases slightly, but as can be seen in Figure \ref{fig: Z dependence}, this effect is almost negligible. $N_{\rm{H}_2}$ increases with metallicity due to the combined effect of increased dust shielding and H$_2$ formation on dust grains. Averaged over time, this effect turns out to be close to linear. Given these effects, the [CII]-to-H$_2$ conversion factor is almost constant with metallicity, with its slight increase attributed to {conversion to C and CO} and the decrease in $f_{\rm{C}^+}$. It is well fitted by the power law $X_{\rm{[CII]}}=6.31\times 10^{19} \;Z^{\prime\;0.17}\; (\rm{cm}^{-2}\;(\rm{K}\;\rm{km}\;\rm{s}^{-1})^{-1})$, which describes our results with an accuracy of $<5\%$. {The middle panel of Figure \ref{fig: Z dependence} compares our computed $X_{\rm{[CII]}}$ to the values found in \cite{Ebagezio2022}, {\cite{Madden2020}, \cite{Vizgan2022}, and \cite{Zanella2018}}. Our [CII]-to-H$_2$ conversion factor is lower than the \cite{Zanella2018} value, and on the low end of the range of values found in \cite{Ebagezio2022}. The bottom panel of Figure \ref{fig: Z dependence} shows the metallicity dependence of $\alpha_{\rm{[CII]}}$, as well as the $W_{\rm{CO}}$ weighted average. The \cite{Zanella2018} value is also presented, but since they compute the H$_2$ mass for a significant fraction of their sample using an $X_{\rm{CO}}\propto Z^{\prime-1.5}$ dependence, we also convert their result so that it is consistent with $X_{\rm{CO}}\propto Z^{\prime-0.71}$ found by \cite{Hu2022a} by multiplying by $Z^{\prime -0.79}$. While our time averaged $\alpha_{\rm{[CII]}}$ is lower than the observed value, it can be explained by some combination of an H$_2$ mass overestimation, especially at low metallicities, and an observational bias towards CO-bright sources. The overestimation of H$_2$ mass is due to the time-dependent effect, discussed in-depth in \cite{Hu2022a}, in which the assumption of chemically steady-state H$_2$ leads to masses larger than found in a model with time-dependent chemistry. This effect is corrected by reproducing the \cite{Zanella2018} value with $X_{\rm{CO}}$ from \cite{Hu2022a}. The observational effect is due to the H$_2$ mass derived from the CO luminosity, i.e., using sources detectable in the CO $J=1-0$ transition only. This drives up $\alpha_{\rm{[CII]}}$ to higher values, as can be seen by comparing our $\alpha_{\rm{[CII]}}$ to the CO-luminosity weighted value. Taking both effects into account, our results fall within a factor of $\sim 3$ of the observed value in all but the $Z^{\prime}=3$ case, a regime which is not well-sampled in \cite{Zanella2018}. A further increase in our computed $\alpha_{\rm{[CII]}}$ might be found by including a more sophisticated model for the ionization state of carbon in HII regions, which could reduce overall [CII] luminosity by $\sim 10-20\%$ (see section \ref{subsec: chemical properties}).}


\begin{figure*}	
	\centering
	\centerline{\includegraphics[trim={0 3.5cm 0 0},clip,width=1.1\linewidth]{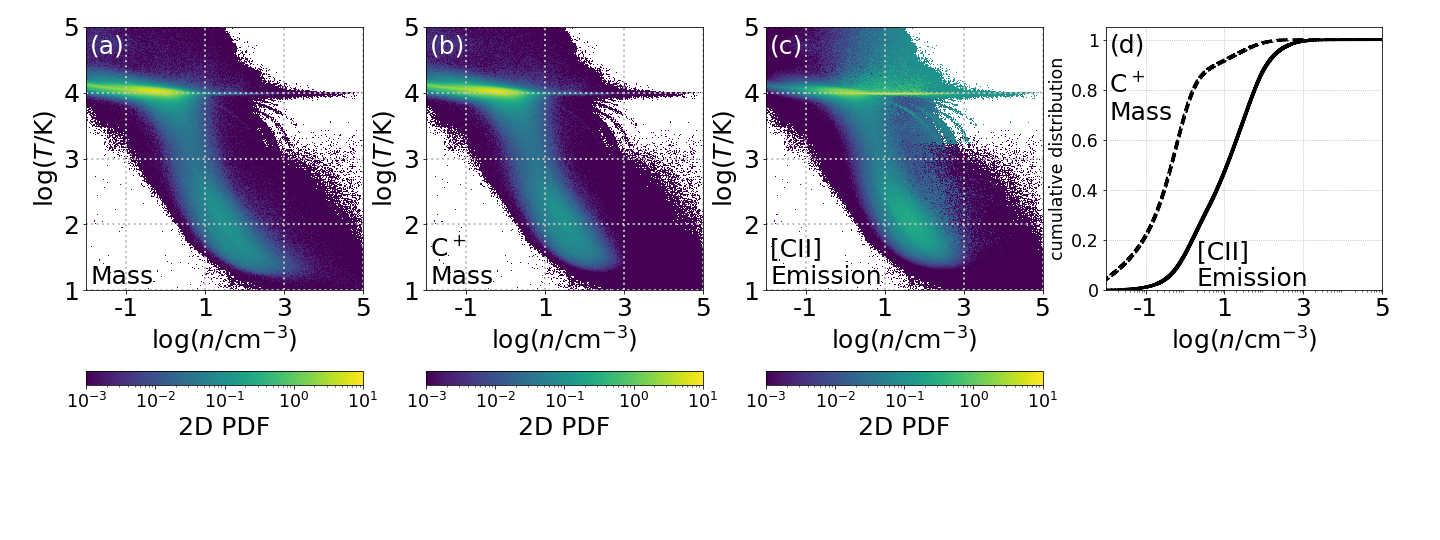}}
	\caption{
2D PDFs in density and temperature weighted by gas mass (panel (a)), C$^+$ mass (panel (b)), and [CII] emissivity (panel (c)). Panel (d) shows the C$^{+}$ mass- and [CII] emissivity-weighted cumulative density distribution.
		}
		\label{fig: 2D histograms}
\end{figure*}

\subsection{Column Density and Emission Maps}

\label{subsec: N and W maps}

Figure \ref{fig: Z=1 snapshots} shows a series of total gas column density maps for the $Z'=1$ run at different times. The life cycle of the ISM is evident. At $t=340$ Myr, the ISM consists of a mostly diffuse medium, with a few small, dense clumps just above and to the right of the image center. Gas cools and accumulates as the clumps grow, until at $t=400$ Myr a dense cloud is fully formed and is visible in the center of the image. As the clouds condense and cool, stars begin to form in the densest regions. The more massive OB stars ionize their surroundings, and once they explode the energy injected into the surrounding gas causes the cloud to disperse, as is visible in the $t=430$ Myr snapshot. This cycle generally repeats and gives a qualitative understanding of the time evolution of the \citetalias{HSvD21} simulations.

Figure \ref{fig: N and W maps} shows maps of H$_2$ and C$^+$ column density, and $W_{\rm{[CII]}}$, for the $Z'=1$ and 0.1 at $t=520$ and 230 Myr respectively. Both snapshots show a dense cloud, traced by large H$_2$ column densities. C$^+$ is more extended, tracing both atomic and molecular gas, while the high column density sightlines trace the molecular gas. 

The $[\rm{CII}]$ emission map is computed using {\sc Radmc-3D}, as described in Section \ref{subsec: RADMC}. The emission intensity qualitatively traces the column density, but with a larger dynamical range, showing that the [CII] intensity is biased towards dense gas. This is due to the high column density sightlines also containing high (volume) density gas, in which collisional excitation of the [CII] line is enhanced. Also visible is the further enhancement of intensity in quasi-spherical bright regions embedded within the dense cloud. These are regions where OB stars photoionize and heat up the gas, causing both an increase in the relative C$^+$ abundance, as well as enhanced excitation of the [CII] line due to the increased temperature and electron fraction. These effects are discussed in detail in Section \ref{subsec: chemical properties}. 

\subsection{Chemical Properties}

\label{subsec: chemical properties}

Since C$^+$ resides in all phases of the ISM, relating [CII] emission to cold gas content and star-formation requires understanding which ISM phase dominates the [CII] emission. Panel (a) in Figure \ref{fig: 2D histograms} shows the 2D histogram for temperature and density weighted by gas mass for the $Z'=1$ run, using all particle data from all snapshots. A few regions of interest are clearly visible. The bright, almost horizontal strip at $\sim10^4$ K extending up to $n\sim10$ cm$^{-3}$ is the WNM, where gas temperature is almost constant across a wide range of densities due to efficient Lyman-$\alpha$ cooling. Above $n\sim10$ cm$^{-3}$, and at lower temperatures, we can see the CNM and molecular ISM, where gas is cooled by metal fine-structure lines (mostly [CII] emission and [CI] 609 and 370 $\mu$m emission) and CO molecular line emission. The thin horizontal strip at $n\gtrsim10$ cm$^{-3}$ is gas in HII regions, where radiation from OB stars ionizes and heats the gas to $10^4$ K. The gas with temperatures higher than the WNM/HII region gas is the hot ionized medium (HIM), shock heated and ionized by SNe feedback. 

We define the different phases in the space of temperature and ionization fraction using the following criteria - cold gas ($T<3\times10^2$ K), unstable neutral medium (UNM; $3\times10^2<T<3\times 10^3$ K), WNM ($3\times10^3<T<3\times 10^4$ K, $x_e<0.99$), HII regions ($3\times10^3<T<3\times 10^4$ K, $x_e>0.99$), and HIM ($T>3\times 10^4$ K). We caution that we may be overproducing the C$^+$ abundance (and therefore [CII] emission) in both HII regions and the HIM, as the carbon is expected to be (at least partially) in the form of higher ionization stages in these environments, while our chemistry treatment does not account for these. In HII regions, a significant fraction of the carbon is expected to be in the form of C$^{2+}$ and C$^{3+}$. As is discussed in Section \ref{subsec: chemical properties}, the overproduction of [CII] in the HIM is not a concern, as the HIM is negligible compared to the other ISM phases. The picture presented in Figure \ref{fig: 2D histograms} is qualitatively similar in simulation runs with different metallicities, the main differences being a lower cold ISM temperature, and a lower transition density between WNM and CNM phases, for higher values of $Z'$.

Panel (b) of Figure \ref{fig: 2D histograms} shows the 2D temperature-density histogram, weighted by C$^+$ mass. We see a similar pattern to that of the total gas mass histogram, the main difference being that C$^+$ mass is absent in the highest density cold gas with $n\gtrsim 10^3$ cm$^{-3}$. This is due to the effect of dust- and gas self-shielding against FUV in the denser regions of gas clouds, reducing the C$^+$ abundance relative to C and CO. Table \ref{table: C+ mass} shows the fraction of C$^+$ mass in the different ISM phases defined above. For all metallicities, most of the C$^+$ mass is in the WNM, where most of the total gas mass is, and where the carbon is entirely in the form of C$^+$. Another fraction of order 10\% is in the cold ISM and UNM across all metallicities. A lower fraction of the mass of the order of a few \% is in HII regions and the HIM. We note that our WNM mass fraction is higher than observed in studies of 21 cm self-absorption \citep{Heiles2003,Stanimirovic2014}.

\begin{table}
\centering
\begin{threeparttable}
\caption{Fractions of C$^+$ mass in different ISM phases.}
\centering 
\begin{tabular}{l l l l l}
\hline \hline 
\\ [-1.5ex] Phase & $Z'=0.1$ & 0.3 & 1 & 3 \\ [0.5ex] 
\hline
\\[-1ex]Cold Gas & 0.038 & 0.055 & 0.112 & 0.142
\\[0.5ex]UNM & 0.088 & 0.057 & 0.033 & 0.063
\\[0.5ex]WNM & 0.844 & 0.850 & 0.855 & 0.746
\\[0.5ex]HII regions & 0.004 & 0.005 & 0.009 & 0.016
\\[0.5ex]HIM & 0.026 & 0.032 & 0.023 & 0.034
\\[0.5ex]
\hline
\end{tabular}
\label{table: C+ mass} 
\end{threeparttable}
\end{table}

The HII region mass fraction is higher for higher metallicity. This can be understood by the fact that the mass of a classical Stromgren sphere around an ionizing source with a given ionizing photon production rate is inversely proportional to gas density. Since the cold ISM cools via line emission from different metal species, it is colder at higher metallicity. This leads to gas reaching the threshold for star formation at lower density, driving down the average density of HII regions as a result. Indeed, the average density of HII regions in our simulations is 28.97, 12.85, 10.06, and 4.61 cm$^{-3}$, for the $Z'=$0.1, 0.3, 1, and 3 runs respectively.

Panel (c) of Figure \ref{fig: 2D histograms} shows the density-temperature histogram weighted by [CII] emissivity. The emissivity of each particle is computed by finding the excitation state of C$^+$ using the density, temperature, and chemical abundances computed in \citetalias{HSvD21}, as well as using collisional rate coefficients from the LAMDA database, as is used by {\sc Radmc-3D}. In this calculation, we assume no radiation trapping, and by using this 3D information regarding [CII] emission we are implicitly assuming that the emission is optically thin. In panel (d) of Figure \ref{fig: 2D histograms} we show the C$^{+}$ mass- and [CII] emissivity-weighted volume density cumulative distribution function, demonstrating the [CII] emission originates in moderately dense gas with $n\sim 10$ cm$^{-3}$, while C$^{+}$ is distributed over a larger density range.

\begin{figure}	
	\centering
	\centerline{\includegraphics[trim={0 0.5cm 0 0},clip,width=1\linewidth]{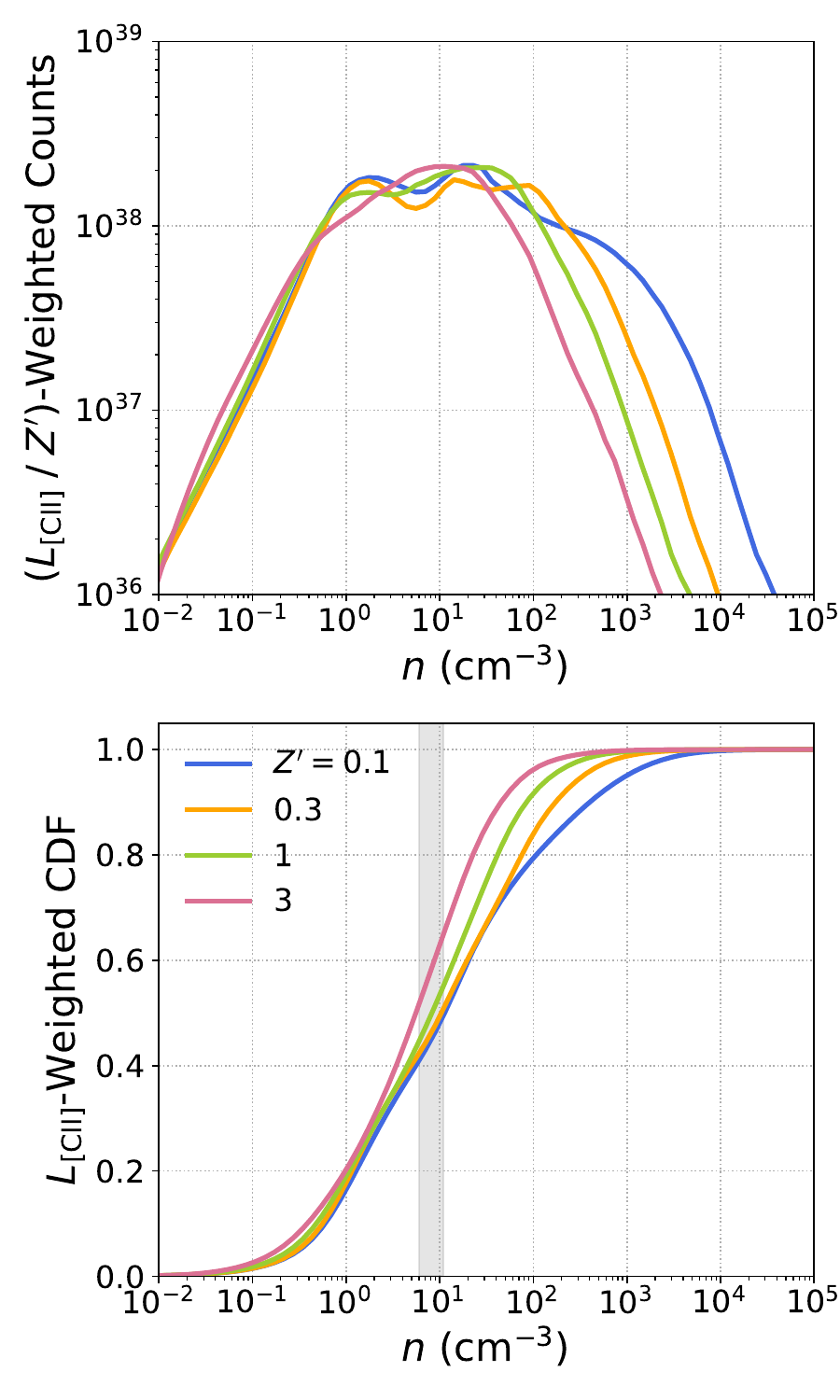}}
	\caption{
Top panel - $L_{\rm{[CII]}}$-weighted density histogram for all for metallicity runs, normalized by metallicity. Bottom panel - $L_{\rm{[CII]}}$-weighted density cumulative distribution functions for all metallicity runs. The range of densities for which the curves reach a value of 0.5 is shown by the shaded grey area.
		}
		\label{fig: small hist}
\end{figure}

{Figure \ref{fig: small hist} shows the $L_{\rm{[CII]}}$-weighted density histogram and cumulative distribution functions for all metallicity runs. The top panel shows that gas with $n\sim10$ cm$^{-3}$ dominates the emission across all metallicities. An increasing contribution from higher-density gas is clearly visible at lower metallicity.
The bottom panel of Figure \ref{fig: small hist} shows that the CDF is only slightly modified, and only at high density where C$^+$ is converted into C and CO, leading to the sublinear $W_{\rm{[CII]}}$-$Z^{\prime}$ relation discussed in section \ref{sec: metallicity dependence overview}.
As the majority of [CII] originates from gas with $n < 100~{\rm cm^{-3}}$,
which is lower than typical densities where the HI-H$_2$ transition occurs,
[CII] mostly traces the cold, atomic (rather than molecular) gas in the ISM.}

As we show in Section \ref{sec: optical depth}, [CII] is only marginally optically thick at $Z'=3$, and optical depth effects are completely negligible at $Z'\leq1$, which validates our discussion of 3D emissivity data. It is clear that while C$^+$ mass resides predominantly in the WNM, [CII] emission originates from cold gas and HII regions, i.e. dense gas. While the fraction of C$^+$ mass in dense gas (cold ISM, UNM, and dense HII regions) is only of order 10\%, the density, being larger by an order of magnitude, enhances the excitation of C$^+$ by an order of magnitude (see Figure \ref{fig: collisional excitation}) and thus increases the emissivity by two orders of magnitude. 






\begin{table}
\centering
\begin{threeparttable}
\caption{Fractions of [CII] luminosity from different ISM phases. The value in brackets shows the result if an ionization correction factor of 0.5 is applied to the HII region contribution.}
\begin{tabularx}{0.47\textwidth}{l l l l l}
\hline \hline 
\\ [-1.5ex] Phase & $Z'=0.1$ & 0.3 & 1 & 3  \\ [0.5ex] 
\hline

\\[-1ex]Cold & 0.19 (0.23) & 0.19 (0.23)& 0.18 (0.23)& 0.18 (0.25)
\\[0.5ex]UNM & 0.10 (0.12) & 0.07 (0.09)& 0.06 (0.08)& 0.06 (0.08)
\\[0.5ex]WNM & 0.35 (0.42)& 0.35 (0.43)& 0.28 (0.36)& 0.18 (0.25)
\\[0.5ex]HII & 0.31 (0.19)& 0.34 (0.21)& 0.44 (0.28)& 0.53 (0.37)
\\[0.5ex]HIM & 0.04 (0.05)& 0.04 (0.05)& 0.04 (0.05)& 0.04 (0.06)
\\[0.5ex]
\hline
\end{tabularx}
\label{table: CII fractions} 
\end{threeparttable}
\end{table}

Table \ref{table: CII fractions} shows the fraction of [CII] emissivity originating in the different ISM phases. It can be seen that while most of the C$^+$ mass is the WNM, its contribution to [CII] emission {is not as dominant, and is in the range of 18-35\%, decreasing with increasing metallicity. Cold gas contributes $\sim20\%$ of emission, almost independent of metallicity. HII regions, while contributing only  $\sim$1\% of the C$^+$ mass, contribute 31-53\% of the emission, increasing with increasing metallicity as HII region density decreases. HII regions are brighter at a given mass than cold gas due to a combination of photoionization heating increasing gas temperature, as well as an increased electron abundance compared to the WNM/CNM. As the rate coefficient with electrons at this temperature is larger than for collisions with H$_2$, this causes an increase in [CII] emissivity by two orders of magnitude relative to cold ISM conditions. Since all carbon gas in HII regions is assumed to be in the form of C$^+$, and its contribution to the total [CII] emission can be dominant, we consider the potential effect of ionization to higher ionization stages of carbon.} \cite{Ebagezio2022} used tabulated CLOUDY models to test the effect of stellar photoionization feedback on the [CII] luminosity in HII regions and find that the [CII] luminosity is reduced by up to 60\% due to ionization of C$^+$ to higher ions. Applying this result to our time-averaged results for [CII] intensity, this would lead to a reduction of between 15 and 26\%, depending on metallicity. {In addition, we ran a sample of CLOUDY models with uniform density for single stars as well as typical star cluster populations in order to understand this potential effect (see Appendix B for a detailed description of our models). We find that for a typical O-B association and gas density typical of HII regions in our simulation it is reduced by $\sim50\%$, broadly consistent with the findings of \cite{Ebagezio2022}. The values in brackets in Table \ref{table: CII fractions} demonstrate the effect of applying an ionization correction factor of 0.5 to the HII region contribution.} HIM contribution is to [CII] is negligible, as expected since HIM is both low mass and diffuse. Using observations of the Milky Way disk, {\cite{Pineda2014} show that $\sim$20\% of [CII] emission originating from ionized gas, while PDRs, cold HI, and H$_2$ account for the remaining $\sim$80\%. This is broadly consistent with our solar metallicity run when applying a correction factor of order $\sim0.5$ to the contribution from HII regions due to higher ionization stages of carbon. \cite{Croxall2017} observe nearby star-forming galaxies with the Herschel Space Observatory, finding that the fraction of [CII] emission originating in ionized gas increases with metallicity, from $\sim10\%$ at $Z^{\prime}\sim0.3$ to $\sim30\%$ at $Z^{\prime}\sim1$. This is again in broad agreement with our ionization corrected results. Comparing to theoretical work proves more difficult, as the choice sub-grid treatment used to determine the gas phase structure strongly affects the result.} 

{Using a similar physical model, \cite{Bisbas2022} simulate the merger of two gas-rich dwarf galaxies with $Z^{\prime}=0.1$, and find that the WNM, CNM, HII regions, and HIM contribute 58, 18, 10, and 14\% respectively. The result is consistent with our findings for the neutral gas, but they find more emission from HIM and less from HII regions. This difference could be due to their different setup, in which the galaxies are gas-rich, as well as experiencing dynamical effects due to the simulated merger. In their cosmological zoom simulation \cite{Olsen2015} found that virtually all [CII] emission originates from PDRs and molecular gas, the latter being more dominant closer to the galaxy center. With a more detailed sub-grid model for [CII] emission, \cite{Olsen2017} found that it is the diffuse ionized gas that dominates [CII] emission. \cite{Vallini2015} found that diffuse CNM contributes up to 40\% of [CII] emission, with the fraction decreasing metallicity or increasing SFR assumed in the sub-grid treatment. \cite{Pallottini2019} found that $\gtrsim 90 \%$ of [CII] emission originates from gas with $n>10\;\rm{cm}^{-3}$. \cite{RamosPadilla2021} applied sub-grid models for [CII] luminosity in a cosmological simulation and found mixed trends with both mass resolution and SFR in the [CII] emission contributions of various ISM phases, with either the diffuse ionized or atomic gas dominating CNM emission.}

{While predictions from theory vary, the contribution of the different ISM phases to [CII] in our results is broadly consistent with observations, yet a more detailed model for carbon ionization could yield an order unity correction to the HII region contribution.}

\begin{figure}	
	\centering
	\includegraphics[width=1\columnwidth]{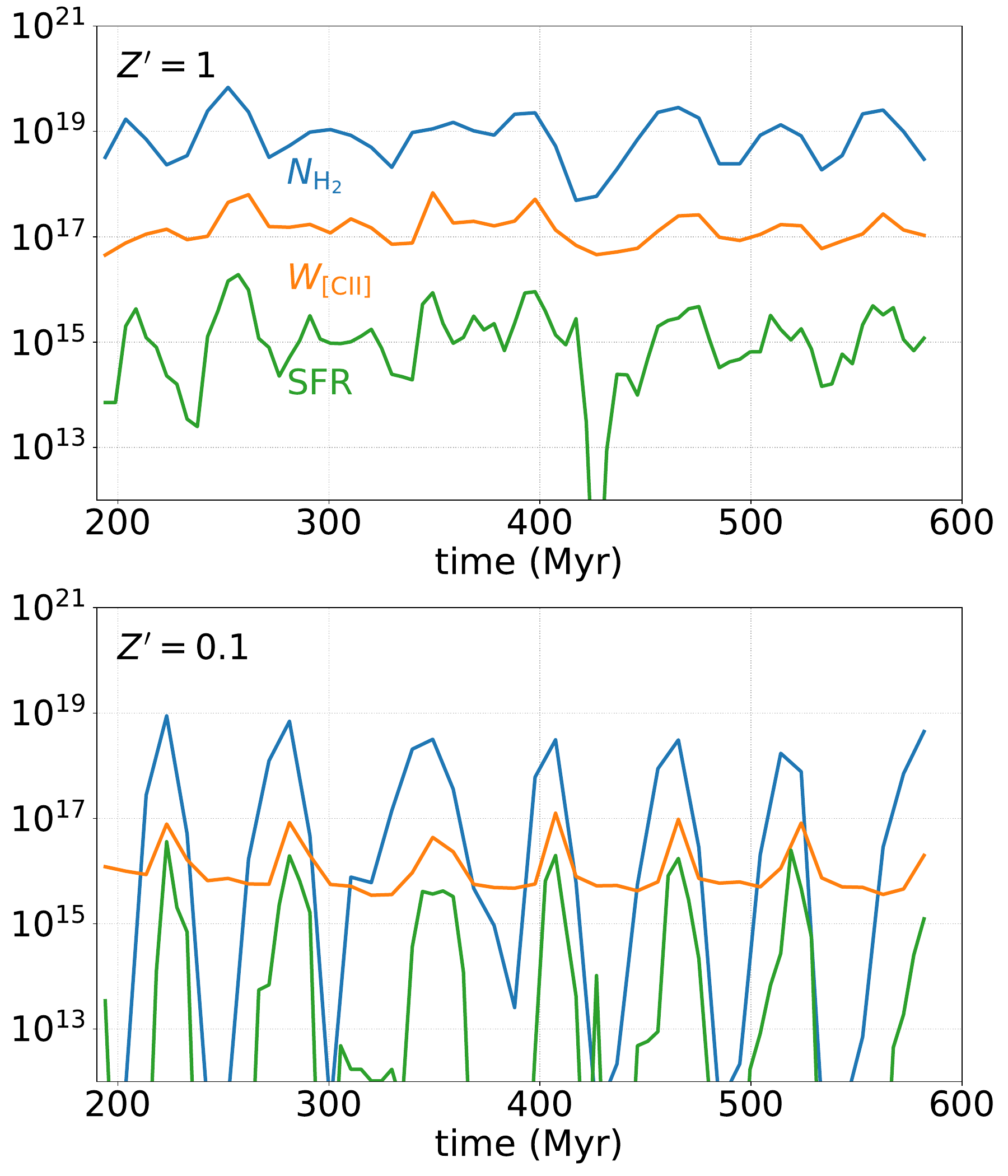} 
	\caption{
Time dependence of SFR/$(10^{-18}\; M_{\odot}\; \rm{yr}^{-1})$, $N_{\rm{H}_2}/(\rm{cm}^{-2})$, and $(W_{\rm{[CII]}}/(10^{-18}\;\rm{K\; km\; s}^{-1}))$, for the $Z'=1$ (top panel) and $Z'=0.1$ (bottom panel) runs.
}
		\label{fig: time dependence}
\end{figure}

\begin{figure}	
	\centering
    \includegraphics[width=1\columnwidth]{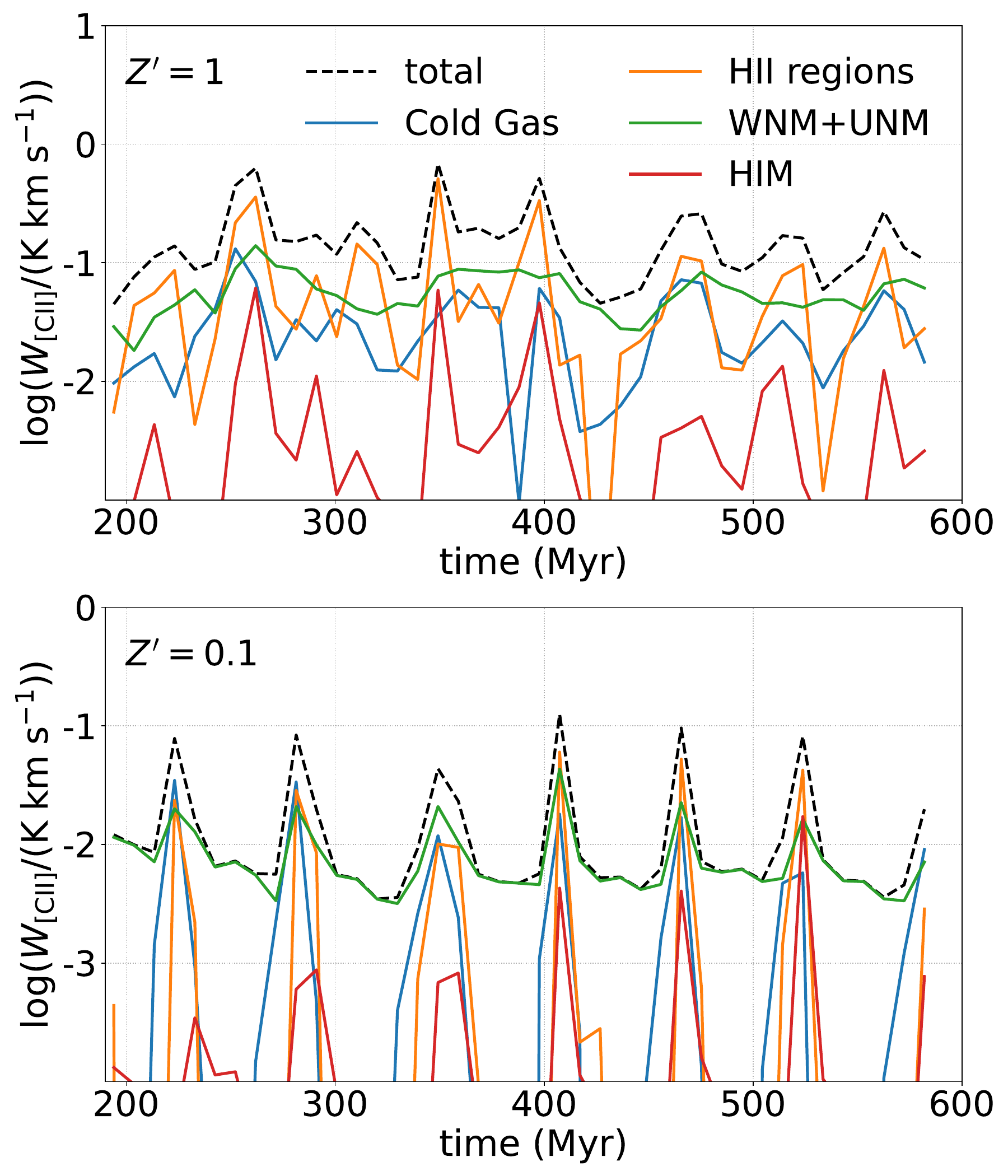} 
	\caption{
Time dependence of [CII] intensity originating from different gas phases: Total intensity, Cold gas (log $T<2.5$), warm neutral medium (including the theramally unstable neutral medium; $2.5<$log$\;T<4.5$, $x_e<0.99$), hot ionized medium (HIM; log $T>$4.5), and HII regions ($3.5<$log $T<4.5$, $x_e>0.99$) for the $Z'=1$ (top panel) and $Z'=0.1$ (bottom panel) runs. 
		}
		\label{fig: emissivity fractions}
\end{figure}

\section{Time Variation}
\label{sec: time var}

In this section we examine the time variation of molecular gas content, star formation rate (SFR), and [CII] emission. The top panel of Figure \ref{fig: time dependence} shows the time dependence of SFR, H$_2$ column density, and [CII] emission, averaged over the entire box at each time step. All three quantities peak at similar times, with SFR falling more dramatically. This is qualitatively understood by a cycle of cloud gravitational collapse followed by stellar feedback, as is demonstrated in Figure \ref{fig: Z=1 snapshots}. The gravitational collapse leads to an enhancement of both density and effective shielding column of the gas. This in turn increases the H$_2$ formation rate and lowers the FUV intensity, leading to high H$_2$ column densities. The density enhancement also increases the [CII] intensity, due to an increased excitation of the upper energy level. The SFR rises as more gas cools and compresses to pass our SF threshold. The rise in SFR triggers photoionization feedback which affects nearby gas. This causes an increase in gas temperature, photodissociation of H$_2$ and CO, and photoionization of C. This leads to a drop in H$_2$ abundance, as well as a short-lived increase in [CII] intensity due to the increased temperature and electron fraction in the HII region, followed by dispersal of the dense gas by SN feedback, which lowers the [CII] intensity. {The effect that the different stellar feedback mechanisms have on [CII] emission have been investigated in different contexts \citep[see, e.g.,][]{Vallini2017}, whereas in this work our sparse time sampling (relative to idealized or analytic models) does not allow for the distinction between the role of SNe and photoionization feedback in setting the time scales for variation in [CII] emissivity.}

This cycle is further demonstrated in the top panel of Figure \ref{fig: emissivity fractions}, where we present the fraction of [CII] emission coming from the different gas phases in the solar metallicity run. The peaks in SFR correspond to peaks in HII region relative emission, while the drops correspond to WNM dominated emission. {Emission from cold gas correlates with HII and is sometimes comparable in magnitude, while} the HIM never contributes significantly to [CII] emission. In comparison, the $Z'=0.1$ run shows different behavior. As is discussed in \citetalias{HSvD21}, the cooling time is longer in lower metallicity gas, leading to burstier and more clustered star formation and, accordingly, long periods of quiescence. This is demonstrated in the bottom panel of Figure \ref{fig: time dependence}, showing the steep drop (and vanishing) SFR between star-formation episodes, while [CII] emission drops much less dramatically to a floor value. During this period of low SFR, the gas remains in predominantly WNM form, leading to reduced emission due to decreased density, but this decrease is not as dramatic as the drop in SFR. This is further demonstrated in the bottom panel of Figure \ref{fig: emissivity fractions}, where cold gas, HII regions, and WNM all contribute to increased emission corresponding to episodes of star-formation, while WNM dominates emission intermittently between the episodes.

\begin{figure*}
	\centering
    \centerline{\includegraphics[width=1\linewidth]{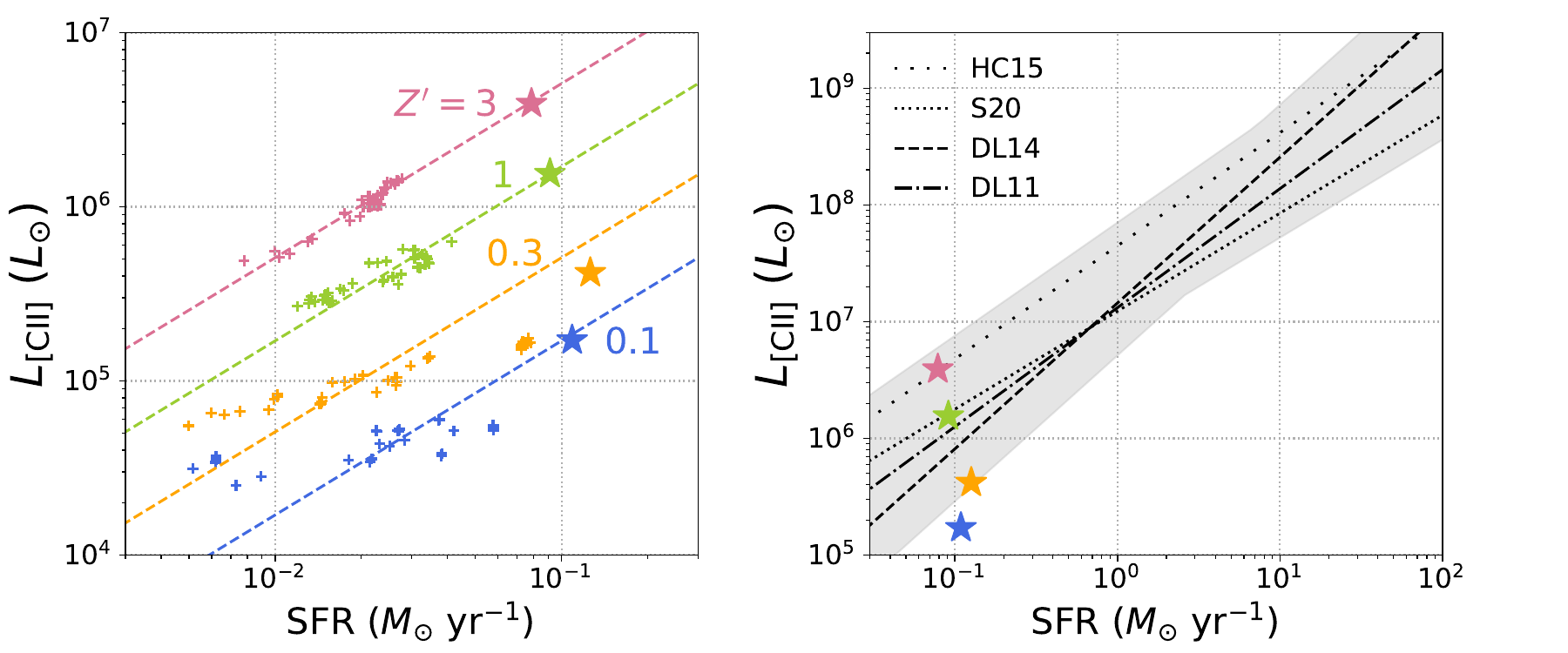}}
	\caption{
Left panel: $L_{\rm{[CII]}}$ vs. SFR, for stacks of 21 snapshots. Stars show stacking of all 41 snapshots for metallicity of the respective color. {Dashed lines are a bilinear fit to $L_{\rm{[CII]}}$ as a function of SFR and $Z^{\prime}$}. Right panel: same stars as for left panel, and fits to observations of \cite{Delooze2011} sample of $z=0$ galaxies, \cite{DeLooze2014} $z=0$ Dwarf Galaxy Survey, \cite{HerreraCamus2015} KINGFISH survey, and \cite{Schaerer2020} ALMA-ALPINE survey of $z\sim4-8$ galaxies. The shaded area represents the union of 1$\sigma$ scatter across all surveys.
		}
		\label{fig: SFR-LCII}
\end{figure*}

\section{$\rm{[CII]}$--SFR Relation}
\subsection{Global $\rm{[CII]}$--SFR Relation}

While a qualitative picture of the relationship between SFR and [CII] emission is established in the previous section, a more quantitative approach is necessary to test theory against observations. To this end, we present in Figure \ref{fig: SFR-LCII} the [CII] luminosity, $L_{\rm{[CII]}}$ vs. star formation rate. $L_{\rm{[CII]}}$ is computed by averaging the line integrated intensity $<I_{\rm{[CII]}}>$, as computed by {\sc Radmc-3D}, over the 1$\times$1 kpc image, and then assuming that the radiation is isotropic, giving
\begin{equation}
    L_{\rm{[CII]}}\equiv 4\pi\times(1~{\rm kpc})^2 <I_{\rm{[CII]}}>.
\end{equation}
Since the relationship between these quantities is usually measured on galactic scales (except for observations of nearby galaxies), we use a stack of galaxy patches, i.e. a sum of $L_{\rm{[CII]}}$ and SFR over 21 snapshots. This is done in order to allow the strong temporal variations in our box to average out as they would on the scale of a galaxy, where we do not expect the ISM across the entire galaxy to be collapsing, forming stars, and dispersing in unison.  We also present the full stack of all 41 snapshots for each metallicity, represented by a star in a corresponding color. Also plotted are fits to observations of different surveys of both local \citep{Delooze2011,DeLooze2014,HerreraCamus2015} and high-redshift galaxies \citep{Schaerer2020}.

A positive trend in the $L_{\rm{[CII]}}$--SFR relation is visible for all metallicities. The scatter is also larger at lower metallicity, due to time variability in these runs being larger than the high metallicity runs. The slope for the $Z'=1$ and 3 runs is close to linear. The $Z'=$0.1 and 0.3 runs show a linear trend at high SFR, but seem to flatten at lower values, {deviating from the linear relation}. This can be explained by the longer cooling time of the low-metallicity runs, which leads to extended periods during which cold, dense gas is absent. This can be seen in the lower panel of Figure \ref{fig: time dependence}, where $W_{\rm{[CII]}}$ hits a floor value when the SFR vanishes. This floor value is associated with WNM emission, as can be seen in the lower panel of Figure \ref{fig: emissivity fractions}. The higher normalization of the different $L_{\rm{[CII]}}$--SFR relation can be understood in two ways. First, as is described in Section \ref{sec: overview}, the total [CII] luminosity increases almost linearly with metallicity, simply due to an increase in the carbon abundance, while the median SFR is constant with metallicity (see \citetalias{HSvD21}). Second, one can understand the increase in the normalization of the relation using an energetics argument. The CNM is heated mainly by dust photoelectric heating, whose abundance scales linearly with metallicity in our simulations, while cooling is dominated by [CII] emission. Heating also scales with FUV intensity, which scales with SFR, so for a constant SFR across all metallicities, we expect [CII] emission to increase linearly to compensate for the increase in photoelectric heating. {Our results can be fit by the expression}
\begin{equation}
    L_{\rm{[CII]}} / L_{\odot} = 10^{7.23}\times Z^{\prime}\times\big{(}{\rm{SFR}}/({M_{\odot}\;\rm{yr}^{-1}})\big{)}
\end{equation}
{with an error $\leq$58\% for SFR $>2\times 10^{-2}\;M_{\odot}\;\rm{yr}^{-1}$. The dashed lines in Figure \ref{fig: SFR-LCII} show the fit for each metallicity run with its respective color.}

In the right panel of Figure \ref{fig: SFR-LCII} we show the stack of all 41 snapshots compared to fits to observational surveys. All but the $Z'=0.1$ data points are within the scatter of the observational data. We mention that our SFR values fall on the low end of the local surveys of \cite{Delooze2011}, \cite{DeLooze2014}, and \cite{HerreraCamus2015}, while the comparison with data from the high-redshift survey of \cite{Schaerer2020} requires an extrapolation of their results, as they do not observe [CII] emission in galaxies with SFR $\lesssim 1 \; M_{\odot}\; \rm{yr}^{-1}$. 

We stress that while stacking our snapshots allows for the averaging out of the temporal scatter in this relation, a more elaborate setup is required in order to fully investigate this relationship on a galactic scale. This is because our box has an average gas surface density corresponding to the solar neighbourhood and a fixed metallicity, while a full galaxy is expected to have a radial variation in both gas surface density and gas metallicity, requiring a stack of many boxes with different gas properties in order to get a picture of the [CII]--SFR on galactic scales. 

\subsection{Resolved {[CII]}--SFR Relation}

{The high ($\sim$ 0.2 pc) spatial resolution of our simulations allows us to compare our results with the spatially resolved [CII]--SFR relation presented in \cite{DeLooze2014} as part of the Dwarf Galaxy Survey \citep{Madden2013}, and \cite{HerreraCamus2015} as part of the KINGFISH project \citep{Kennicutt2011}, both carried out using the Herschel Space Observatory. For a proper comparison, we first use a Gaussian filter with a FWHM of 250 pc, the median physical beam size for the resolved observations presented in \cite{DeLooze2014}. The median resolution of the \cite{HerreraCamus2015} sample is 650 pc. We then plot the median and $\pm25\%$ range over all pixels in all snapshots of our simulation, presented In Figure \ref{fig: Resolved SFR-LCII}, as well as the best fit to observations. Our results show a metallicity trend towards stronger [CII] emission for a given SFR, as is seen in the global [CII]--SFR relation, and is explained in the same way. The metallicity range of the \cite{DeLooze2014} sample is 0.1-0.38. Our $Z^{\prime}=0.3$ results fall within the scatter of their data, except for a low number of points at the high $\Sigma_{\rm{[CII]}}$ end. Our $Z^{\prime}=0.1$ is visibly offset from their relation. This could be due to the metallicity inhomogeneity in their sample, and is discussed further in \ref{subsec: metallicity independence}. The mean metallicity if the \cite{HerreraCamus2015} sample is 0.5. It is consistent with our results for solar metallicity, but shows a higher [CII]--SFR ratio than our $Z^{\prime}=0.3$ run.}

\subsection{Metallicity Dependence of The [CII]--SFR Relation}
\label{subsec: metallicity independence}

\begin{figure}	
	\centering
    \centerline{\includegraphics[width=1\columnwidth]{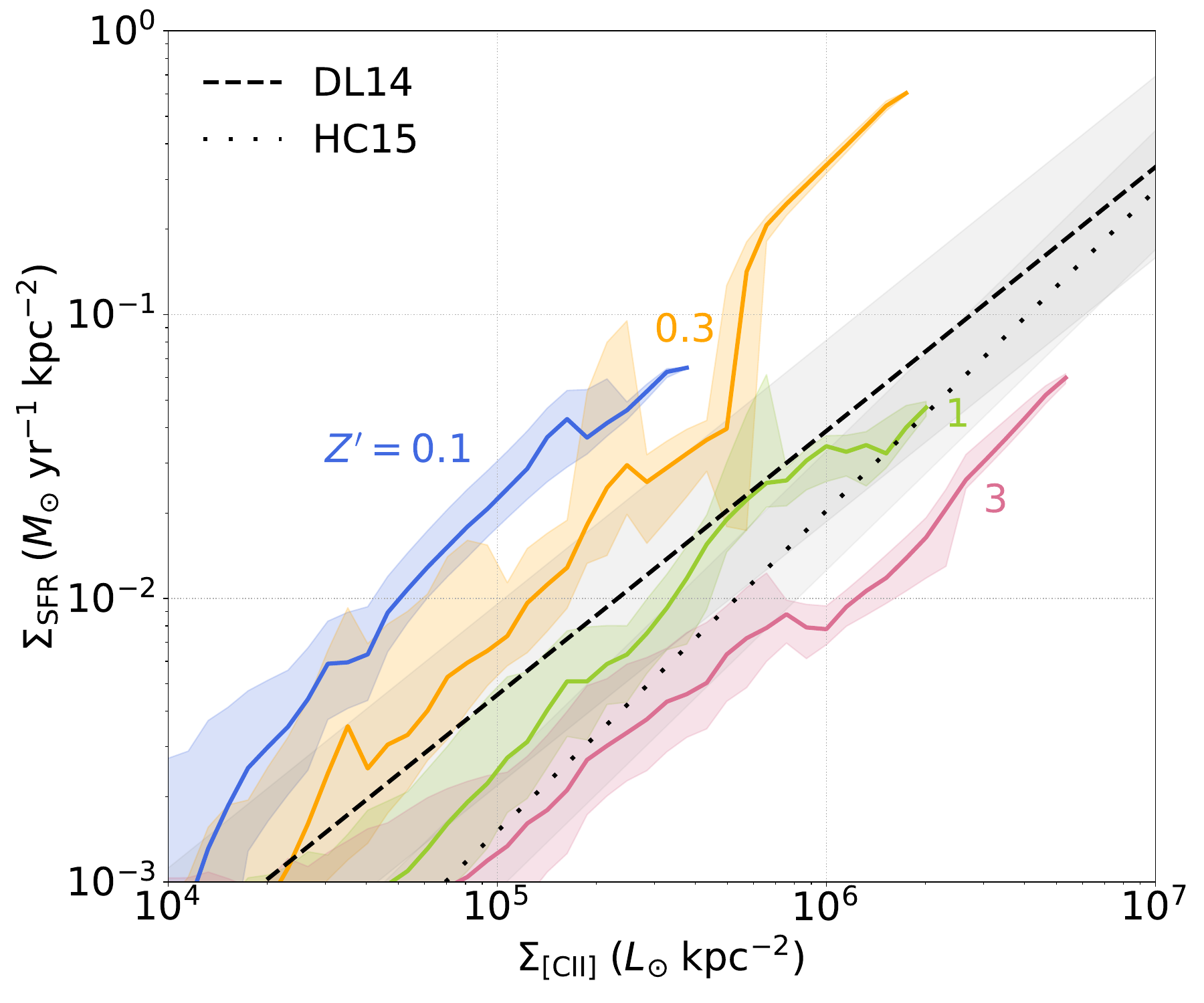}}
	\caption{Median values for our resolved [CII]-SFR relation, obtained using a Gaussian filter with FWHM of 250 pc and averaging all pixels from all snapshots of each run. Shaded regions show the $\pm25\%$ range in the data. Also shown are the \cite{DeLooze2014} and \cite{HerreraCamus2015} best fit for their resolved galaxy samples, with $1\sigma$ scatter in shaded grey.
		}
		\label{fig: Resolved SFR-LCII}
\end{figure}

\begin{figure}
	\centering
    \centerline{\includegraphics[width=1\linewidth]{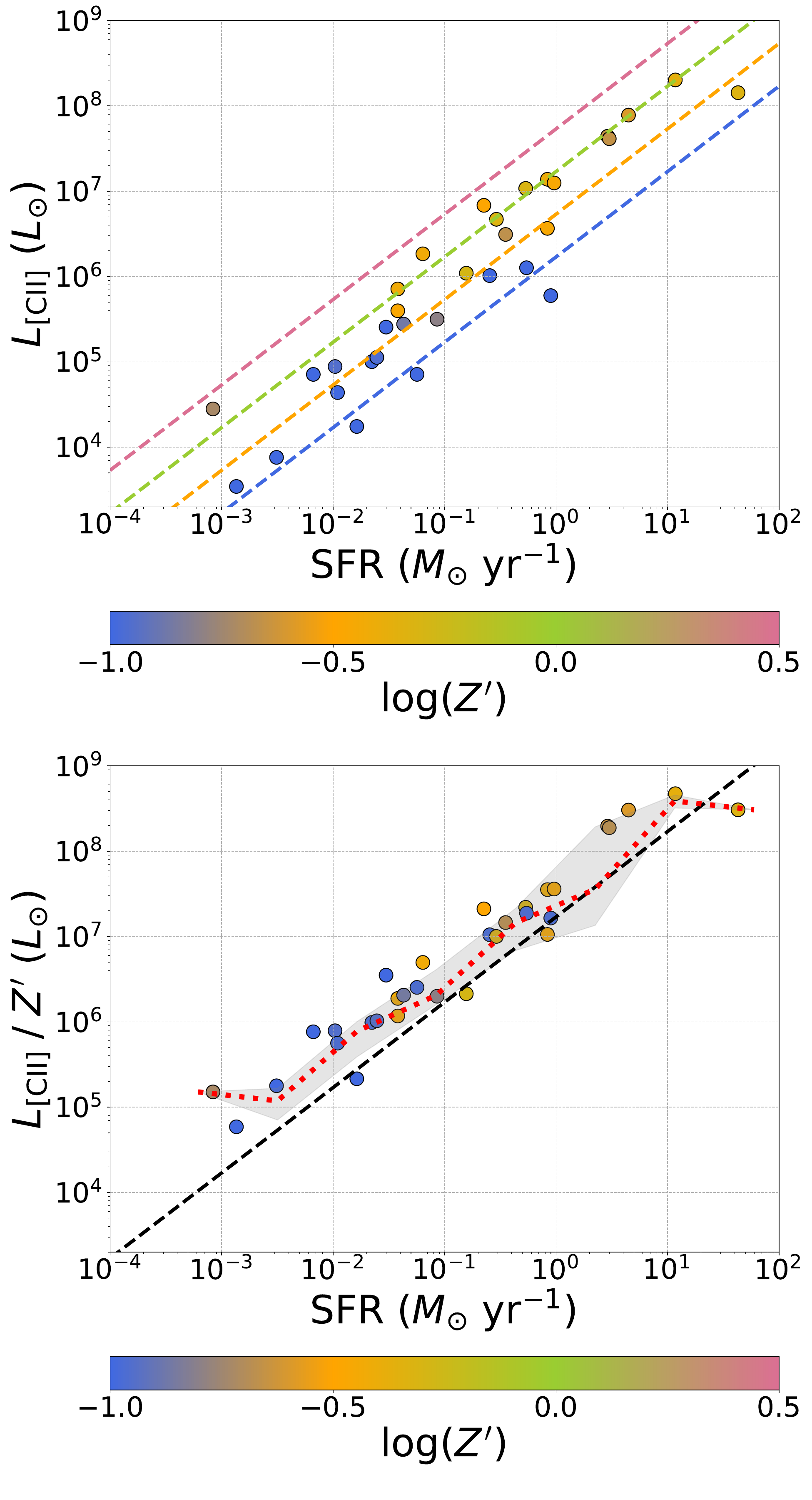}}
	\caption{
Top panel: $L_{\rm{[CII]}}$ vs. SFR from \cite{DeLooze2014}. Data points are colored by metallicity. Dashed lines are the fit to our simulation data for SFR$>2\times10^{-2}\;M_{\odot}\;\rm{yr}^{-1}$, colored similarly. Bottom panel: as for the top panel, but with $L_{\rm{[CII]}}$ normalized by metallicity. The dashed black line is the fit to our solar metallicity data. The red dashed line is the median of observed $L_{\rm{[CII]}}/Z^{\prime}$ values, and the shaded region corresponds to $\pm25\%$.
		}
		\label{fig: SFR-LCII Z dep}
\end{figure}

{
Our simulations 
show a clear metallicity dependence of the [CII]--SFR relation. While there is some evidence for this relation at $z\sim0$, there is no clear consensus on the matter. Figure \ref{fig: SFR-LCII Z dep} shows the data from \cite{DeLooze2014} plotted against the fit to our results. The observations are colored by metallicity and show a qualitative positive trend in their [CII]--SFR ratio. The bottom panel of Figure \ref{fig: SFR-LCII Z dep} shows the same galaxies with their [CII] luminosity scaled down with metallicity, plotted against our solar metallicity fit. Normalizing by metallicity decreases the scatter in the observed [CII]--SFR relation, and shows agreement with our fit to within a factor of $\sim$2. This is preliminary evidence of our results successfully reproducing the observed metallicity trend. A larger sample and a larger range of metallicities would allow for a statistical test of the metallicity dependence. Given our findings, we suggest that the agreement between our results and observations presented in Figure \ref{fig: Resolved SFR-LCII} would be improved by using more metallicity-homogeneous galaxy samples.}

{\cite{Liang2023} claimed that the metallicity-independent [CII]--SFR relation at low redshift is due to a decrease in the relative abundance of C$^+$, as dust shielding against FUV leads to more efficient conversion of carbon to C and CO, which in turn balances the increase in total carbon abundance with increasing metallicity. We do not observe this trend, as is demonstrated by Figure \ref{fig: small hist}, where [CII] emission is dominated by moderately dense gas with $n\sim10$ cm$^{-3}$ across all metallicities, a density which is lower than the typical transition density from C$^+$ to C. This difference could arise from the difference in the implementation of FUV shielding. While we use a HealPIX-based method to estimate the effective shielding column for every particle in the simulation, accounting for the effects of the porosity of the ISM, \cite{Liang2023} consider each gas particle (with a mass of $\sim 10^4\;M_{\odot}$) as a perfect sphere, which could result in overestimated shielding columns and consequently a higher relative abundance of C and CO.}

\section{Optical Depth}

\label{sec: optical depth}

\subsection{Optically Thin Calculation}

\begin{figure}
	
	\centering
    \centerline{\includegraphics[width=1.2\linewidth]{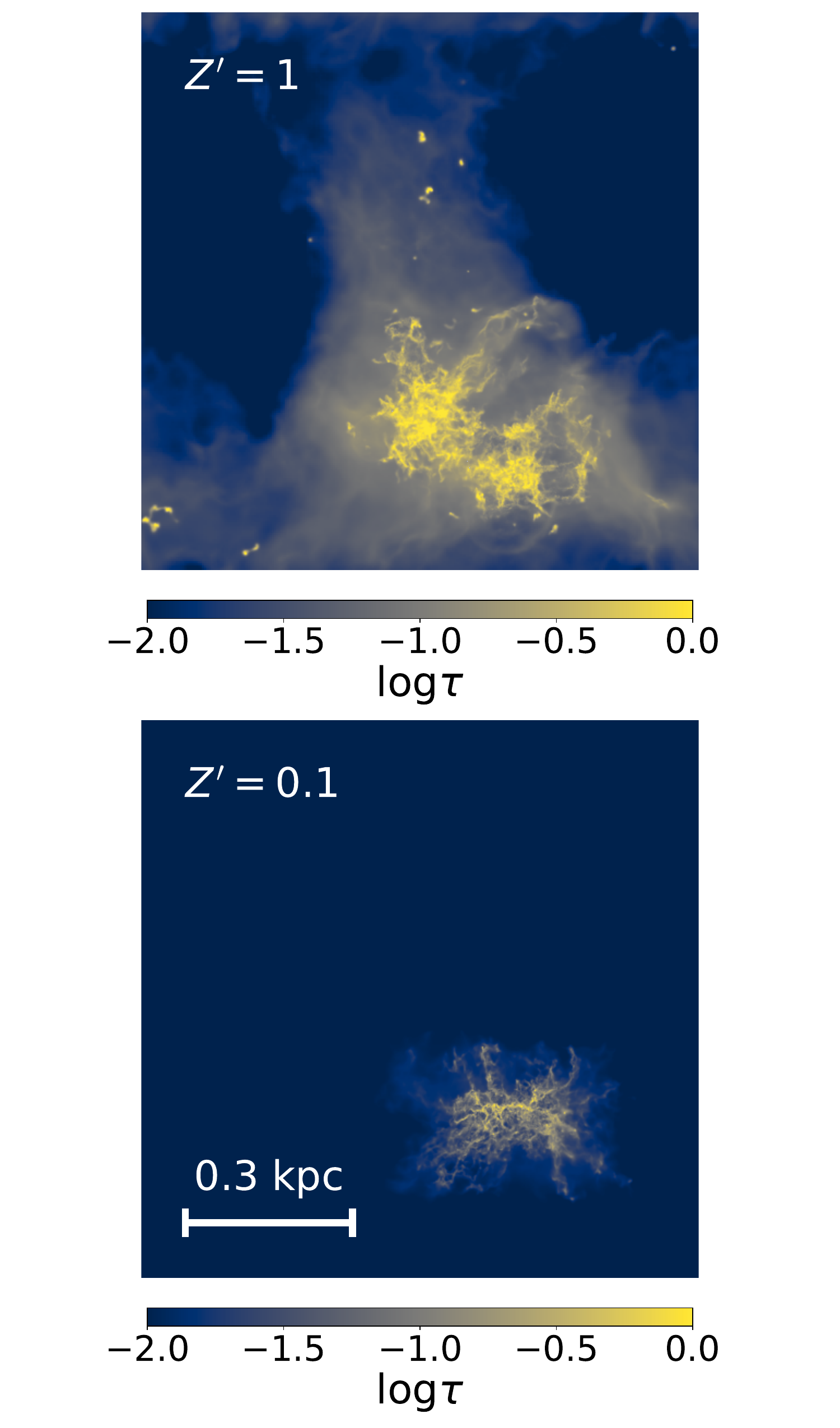}}
	\caption{
Maps of projected [CII] optical depth for the $Z'=1$ (top panel) and 0.1 (bottom panel), for $t=520$ and 230 Myr respectively.
		}
		\label{fig: optically thin map}
\end{figure}

To investigate the effects of optical depth on our results, and validate our numerical computations, we compare the results computed using {\sc Radmc-3D} with an independent optically thin calculation. We compute the [CII] emissivity for each cell in the AMR grid, by neglecting absorption and stimulated emission, and using the temperature, density, and chemical abundances calculated in the simulation, as well as the rate coefficients from the LAMDA database. The emissivity is given by

\begin{equation}
    j_{10}=n_{1}A_{10}E_{10},
\end{equation}
where $n_1$ is the density of excited C$^+$, $A_{10}$ is the Einstein coefficient for spontaneous emission, and $E_{10}$ is the transition energy. For a two-level system such as C$^+$, $n_1$ can be calculated simply using

\begin{equation}
    \frac{n_1}{n_0}=\frac{R_{01}}{R_{10}}=\frac{\sum_i k_{01,i}n_i}{\sum_i k_{10,i}n_i+A_{10}},
\end{equation}

where $n_0$ is the density of the lower excitation state of C$^+$, $n_i$ is the density of collision partner $i$, and $k_{01,i}$ ($k_{10,i}$) is the upwards (downwards) collisional rate coefficient with collision partner $i$. $R_{ij}$ is the rate of transitions from level $i$ to level $j$.

We then project the emissivities onto a 512x512 grid, effectively creating an optically thin image of the line integrated intensity. 

In addition, we compute the optical depth for each cell given by 

\begin{equation}
    \frac{d\tau_{10}}{dz} = \frac{\lambda_{10}^3A_{10}g_1}{8\pi^{3/2}g_0} \Big{(}\frac{n_0}{b}(1-\frac{n_1g_0}{n_0g_1})\Big{)},
\end{equation}
where $\lambda_{10}$ is the wavelength of the [CII] line, $g_1$ ($g_0$) is the statistical weight of the upper (lower) excitation stat of C$^+$, and $b$ is the Doppler broadening parameter. $b$ is calculated using
\begin{equation}
    b^2=b_{\rm{turb}}^2+b_{\rm{therm}}^2=b_{\rm{turb}}^2+\frac{k_BT}{\mu m_p},
\end{equation}
where $b_{\rm{turb}}$ and $b_{\rm{therm}}$ are the turbulent and thermal broadening parameters, respectively, $k_B$ is the Boltzmann constant, $\mu$ is the mean molecular weight, and $m_p$ is the proton mass. We then project $d\tau_{10}/{dz}$ onto the same grid and create a map of the optical depth $\tau_{10}$. In Figure \ref{fig: optically thin map} we show sample optical depth maps for the snapshots presented and discussed in Section \ref{subsec: N and W maps}. In the $Z'=1$ case, $\tau_{10}\ll 1$ for sightlines outside of the dense cloud. In the densest regions, however, $\tau_{10}$ reaches unity. In the $Z'=0.1$ case, $\tau_{10}$ is small everywhere. Our computation of the optical depth is an upper limit, and in fact, is an overestimation in the high $W_{\rm{[CII]}}$ sightlines by $\sim$0.5 dex. The origin of this overestimation is due to {\sc Radmc-3D} taking into account the bulk velocity of the gas particles, which can allow radiation to escape freely through a high column density sightline. Discussion of this effect, as well as a more detailed comparison between our optically thin calculation and {\sc Radmc-3D}, can be found in Appendix A. The comparison shows a good agreement between {\sc Radmc-3D} and our optically thin calculation in the optically thin sightlines, and the difference between the two in the optically thick regime can be explained by the optical depth calculation.

\begin{figure}
	
	\centering
	\centerline{\includegraphics[width=1.1\linewidth]{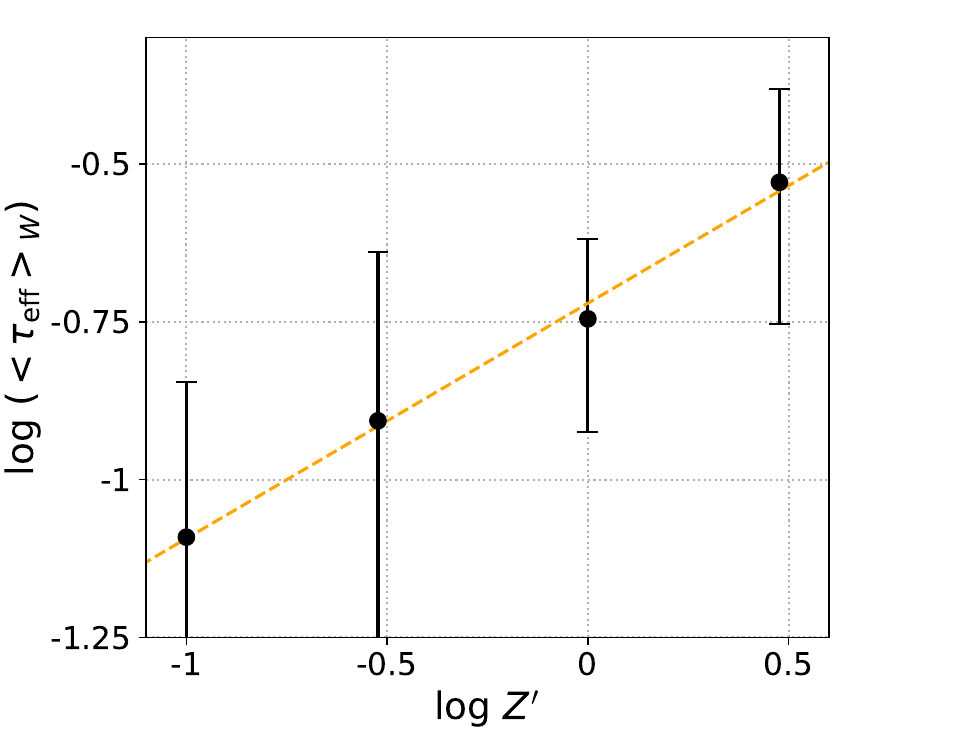}}
	\caption{
Effective intensity-weighted [CII] optical depth averaged over the entire simulation run. as a function of gas metallicity. Error bars represent the standard deviation of the 41 snapshots averaged for each simulation run.
		}
		\label{fig: tau vs Z}
\end{figure}

Given the agreement between our optically thin calculation and the full radiative transfer, we can justify our calculations in Section \ref{subsec: chemical properties} based on the computed 3D emissivity. In addition, we can now investigate the optical depth of [CII] emission in our simulation.

\subsection{Optical Depth Effect}
Given that we overestimate optical depth at the high end by approximately 0.5 dex, we opt for a more accurate estimation of optical depth. We define the effective optical depth by the equation
\begin{equation}
    \frac{1-e^{-\tau_{\rm{eff}}}}{\tau_{\rm{{eff}}}}=\frac{W_{\rm{[CII],RADMC-3D}}}{W_{\rm{[CII],thin}}},
    \label{eq: tau eff def}
\end{equation}
where $W_{\rm{[CII],RADMC-3D}}$ and $W_{\rm{[CII],thin}}$ are the values for $W_{\rm{[CII]}}$ in our full radiative transfer and optically thin calculation, respectively. This definition is motivated by the solution of the radiative transfer equation for a static, uniform medium with source function $S$, no background radiation, and optical depth $\tau$. The full solution results in an intensity
\begin{equation}
    I=S(1-e^{-\tau}),
\end{equation}
while the solution for the optically thin limit gives
\begin{equation}
    I_{\rm{thin}}=S\tau,
\end{equation}
and the ratio between the two solutions is given by the expression in Equation \ref{eq: tau eff def}, replacing $\tau_{\rm{eff}}$ with $\tau$.

For each simulation snapshot, we then compute the intensity weighted average over all pixels in the snapshot
\begin{equation}
    <\tau_{\rm{eff}}>_W\equiv \frac{\sum W_i\tau_{\rm{eff},i}}{\sum W_i}.
\end{equation}
Figure \ref{fig: tau vs Z} shows the time average of $<\tau_{\rm{eff}}>_W$ as a function of metallicity. Importantly, we find that for the gas surface density in our simulations, the gas only just approaches the regime where optical depth effects are important at $Z^{\prime}=3$, where the time-averaged effective optical depth gives an optically thin intensity stronger by only ~15\%. We present a power law fit to our data of the form
\begin{equation}
    \mathrm{log} (<\tau_{\mathrm{eff}}>)=0.37 \times \mathrm{log} Z^{\prime}-0.72.
\end{equation}
This gives the result that [CII] emission can be assumed to be (approximately) optically thin up to solar metallicity (at all metallicities), given a gas surface density of 10 $M_{\odot}$ pc$^{-2}$. It is highly probable that this result will not hold in environments where gas surface densities are higher by an order of magnitude or more, which should scale the optical depth by the same factor.

\section{Summary}
\label{sec: summary}

We have presented radiative transfer calculations of [CII] intensity using {\sc Radmc-3D} and applied it to a set of sub-pc resolution hydrodynamical simulations of a self-regulated, SN-driven ISM. We investigate the systematic effects of metallicity on line intensity over a large range $0.1\leq Z'\leq 3$, and the intensity contribution of different ISM phases. Our simulation time (500 Myr) and volume ($1\times1\times10$ kpc$^3$) are large enough to allow us to understand the spatial and temporal variations in [CII] intensity and its correlation with SFR and molecular gas content. We summarize our main results below:

\begin{enumerate}[leftmargin=*, noitemsep]
    
    \item The time-averaged [CII] line luminosity on kpc scales grows approximately linearly with metallicity at all but super-solar metallicities (see Figure \ref{fig: Z dependence}). This effect is simply a column density effect, as most of the carbon mass is in the form of C${^+}$, and the emission is approximately optically thin, especially in the sub-solar metallicity cases. 
    
    \item The spatially averaged [CII] luminosity correlates with SFR, as long as it is dominated by {dense gas, with a potential contribution from HII region, cold gas, and WNM, depending on metallicity. This generally holds for for SFR $>2\times 10^{-2}\;M_{\odot}\;\rm{yr}^{-1}$}. We find a linear $L_{\rm{[CII]}}$--SFR relation, with a normalization proportional to metallicity (see left panel of Figure \ref{fig: SFR-LCII}). {For SFR $>10^{-2}\;M_{\odot}\;\rm{yr}^{-1}$ it is can be described by the fit} 
    \begin{equation*}
        L_{\rm{[CII]}} / L_{\odot} = 10^{7.23}\times Z^{\prime}\times\big{(}{\rm{SFR}}/({M_{\odot}\;\rm{yr}^{-1}})\big{)}
    \end{equation*}
    {within an error $\leq58\%$.} This is due to both dust photoelectric heating and C$^+$ emissivity scaling linearly with metallicity. This is also explained by SFR correlating with the gravitational collapse and clumping of gas, which in turn enhances collisional excitation and, in turn, emissivity.
    {Our results fall within the large scatter of observations of the [CII]--SFR relation, and normalizing $L_{\rm{[CII]}}$ by $Z^{\prime}$ improves the agreement with the $z\sim0$ samples and provides evidence for a metallicity dependence of the [CII]--SFR relation.}
    
    \item Most of the C$^+$ mass is in the WNM, but [CII] emission is skewed towards denser gas, with the cold gas and dense HII regions contributing significantly. The majority of [CII] emission originates from gas with $n>10$ cm$^{-3}$, i.e., dense but not necessarily molecular gas. This also highlights the need to resolve the small-scale clumpy structure of the ISM to realistically capture the [CII] line.
    
    \item We find that different phases of the ISM dominate emission at different times (see Figure \ref{fig: emissivity fractions}, Table \ref{table: CII fractions}). Over the entire simulation run, it is the WNM and HII regions that dominate emission, the latter's higher density enhancing the excitation of the [CII] line. Lower luminosities associated with the lowest SFRs ($\lesssim10^{-2}\; M_{\odot }$ yr$^{-1}$) represent a floor value that is dominated by WNM emission, and no longer correlates with SFR. This regime is only explored in the sub-solar metallicity runs, as longer cooling times lead to extended periods in which dense gas is absent. We caution that our treatment of emission from HII regions does not include higher ionization stages of carbon and emission from HII regions is overproduced by up to a factor of $\sim$ 2.
    
    \item  {[}CII] emission traces H$_2$ (see Figure \ref{fig: Z dependence}), for the same reasons that it traces SFR. The metallicity dependence of the conversion factor $X_{[\rm{CII]}}$ is weak, ranging from 5 to 9$\times 10^{19}$ cm$^{-2}$/(K km s$^{-1}$). This is due to [CII] luminosity and H$_2$ column density having a similar scaling with metallicity. In conjunction with our findings that [CII] emission is dominated by dense gas, this is evidence in support of using [CII] as a tracer for molecular gas content. Our results are fit well by the power law $X_{\rm{[CII]}}=6.31\times 10^{19} \;Z^{\prime\;0.17}\; (\rm{cm}^{-2}\;(\rm{K}\;\rm{km}\;\rm{s}^{-1})^{-1})$. For solar metallicity, we find agreement between our results and \cite{Ebagezio2022}. {Our computed value for $\alpha_{\rm{[CII]}}$ is lower than the observed value in \cite{Zanella2018}. When accounting for non-equilibrium effects on H$_2$ formation and taking into account potential observational bias towards CO-bright sources, the tension between the two results is partially relieved, albeit not at $Z^{\prime}=3$, which is poorly sampled by the observations.} 
    
    \item By carefully comparing a full radiative transfer treatment with our own optically thin calculation, we are able to quantify the effective optical depth, accounting for absorption, relative bulk motion, and line-broadening (see Figure \ref{fig: tau vs Z}). We find that for average gas surface density of $\Sigma=10\;M_{\odot}$ pc$^{-2}$, [CII] emission can be treated as optically thin except for the most extreme cases in our $Z'=3$ run, where $\tau\sim 0.3$.
\end{enumerate}

We thank Chris McKee, Reinhard Genzel, and Ulrich Steinwandel for fruitful discussions. We thank the referee for their very helpful and constructive comments. This work was supported by the German Science Foundation via DFG/DIP grant STE/1869-2 GE 625/17-1, by the Center for Computational Astrophysics (CCA) of the Flatiron Institute, and the Mathematics and Physical Sciences (MPS) division of the Simons Foundation, USA.

\bibliography{library}
\bibliographystyle{aasjournal}

\section*{Appendix A - Comparison Between {\sc Radmc-3D} and The Optically Thin Calculation}

\label{Appendix A}

\begin{figure*}
	
	\centering
	\includegraphics[width=1\columnwidth]{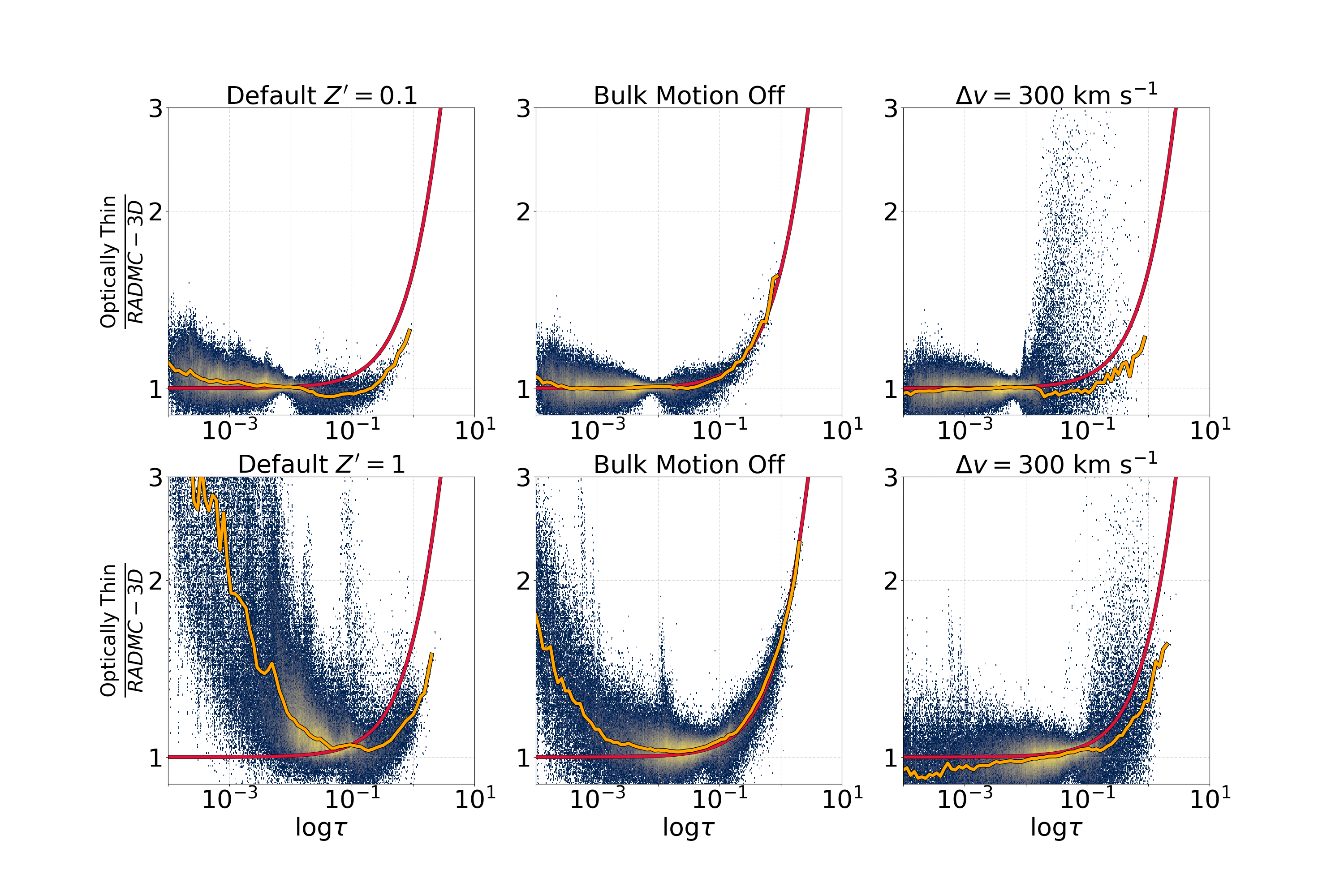} 
	\caption{
Pixel-by-pixel ratio between our optically thin calculation and the full {\sc Radmc-3D} calculation as a function of projected optical depth for our default settings (left panel), the linewidth set to 200 km s$^{-1}$ (middle panel), and with gas velocity set to 0 (right panel). The green curve shows the value theoretical value computed for a uniform gas mix.
		}
		\label{fig: optically thin scatter comp}
\end{figure*}

In this appendix, we describe in detail our comparison between the optically thin calculation and {\sc Radmc-3D}. Figure \ref{fig: optically thin scatter comp} shows the ratio between the {\sc Radmc-3D} calculation and the optically thin calculation as a function of optical depth, for the $Z'=1$ and 0.1 at $t=520$ and 230 Myr respectively, using three different methods. The left column shows the result using our default settings on {\sc Radmc-3D}, as described in Section \ref{subsec: RADMC}. The middle column shows the same calculation as (a) but with the bulk velocity of all particles set to 0, and {\sc Radmc-3D} line transfer mode set to "optically thin". We stress that this line transfer mode does not ignore absorption, but the "optically thin" term here refers to the calculation of the local level populations assuming no line trapping, as opposed to our default settings where line trapping is taken into account using the LVG approximation. The right column shows the same calculation, but this time the wavelength range covered by {\sc Radmc-3D} when integrating over the [CII] line was set to 300 instead of 20 km s$^{-1}$, and again using "optically thin" line transfer mode in {\sc Radmc-3D}. All panels show the curve 
\begin{equation}
    \frac{I_{\rm{thin}}}{I_{\rm{full RT}}} = \frac{\tau}{1-e^{-\tau}},
\end{equation}
as is the case for a uniform slab. The default settings show good agreement at intermediate optical depth, but an overestimation of optically thin emission at both low and high optical depth. The middle column shows that when the bulk velocity of the gas is set to 0, the overestimation of optical depth in the optically thin calculation is mitigated. This shows that the default comparison suffers from an overestimation of optical depth, due to neglecting the relative bulk motion of the gas in the optically thin calculation. The turning off of LVG mode here was in order to avoid numerical problems that arise from computing the velocity gradient when the velocities are set to 0 everywhere. The right column shows that the disagreement at low optical depth is a result of the low optical depth sightlines containing gas with a large linewidth, as well as possibly large bulk motions. This causes a significant fraction of the emission to be missed by {\sc Radmc-3D} during integration, as line emission is either broadened or shifted out of the integration window, and to an underestimation of emission. Since these low optical depth sightlines are made up of very low density and low column density gas, their intensity is low and this difference does not affect our results in a significant way when averaged over an entire snapshot. Overall, our optically thin calculation reproduces the {\sc Radmc-3D} calculation well, and overproduces emission to the expected extent.

\section*{Appendix B - [CII] Emission From HII Regions}

\label{Appendix B}

\begin{figure*}
	
	\centering
	\includegraphics[width=1\columnwidth]{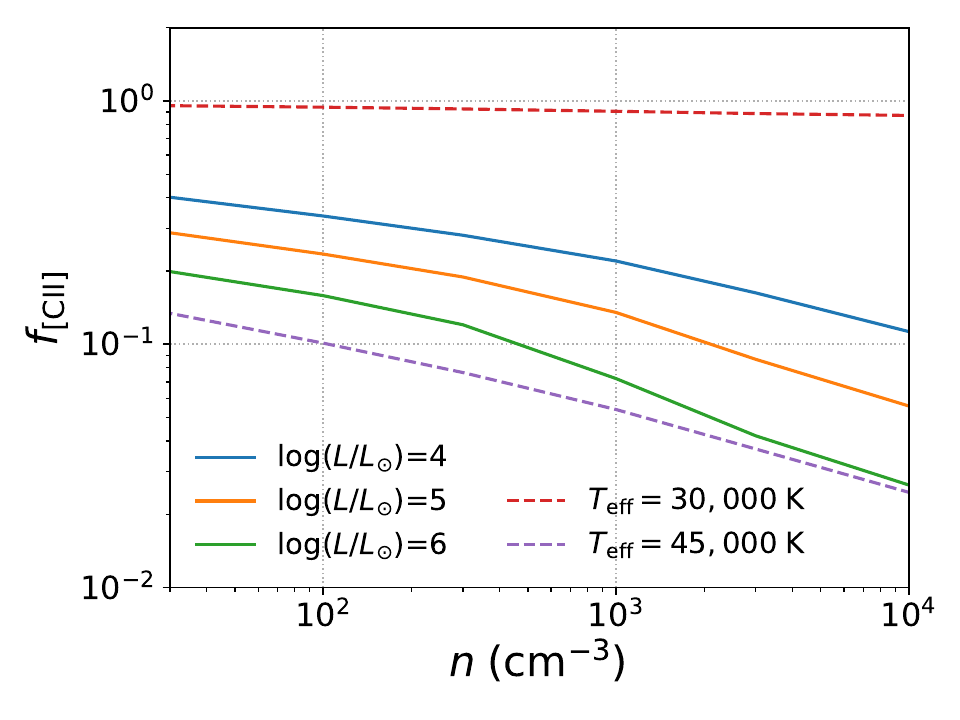} 
	\caption{
$f_{\rm{[CII]}}$ as a function of gas density for our CLOUDY models. Solid lines show the results for radiation sources with a spectral shape calculated using STARBURST99. Dashed lines are for single main-sequence O-B stars from the stellar atmosphere library of \cite{Castelli2003}.
		}
		\label{fig: CLOUDY model}
\end{figure*}

{In this appendix we present a simple grid of CLOUDY \citep{Ferland2017} models to demonstrate the potential effect of ionizing radiation from massive stars on [CII] emission. As is presented in section \ref{subsec: chemical properties}, [CII] emission from HII regions contributes a significant fraction of total emission, and dominates in the $Z^{\prime}=1$ and 3 runs. This is not the case for observations of the Milky Way and nearby galaxies, and is due to our assumption that carbon in HII regions is entirely in the form of C$^+$. Observed HII can have a large fraction of carbon in higher ionization stages \citep{Draine2011}, but the effect is dependent on both the radiation source and gas conditions. \cite{Ebagezio2022} found that including a more detailed treatment of carbon ionization reduced [CII] emission by up to $\sim60\%$. For our photoionization models, we set up a constant density, spherical gas cloud, with solar heavy element abundances, illuminated by a central radiation source. The shape of the source spectrum is assumed to be either a single star \cite{Castelli2003} stellar atmosphere model, or calculated using the evolutionary stellar population synthesis code STARBURST99 \citep{Leitherer1999}, assuming a \cite{Kroupa2002} initial mass function, star formation rate of $1\; M_{\odot}\;\rm{yr}^{-1}$, and a lower (upper) mass cutoff of 0.1 (100) $M_{\odot}$, and an age of 100 Myr. For each CLOUDY model, the output includes the relative abundance of C$^{+}$ as a function of radial distance from the radiation source, which we then integrate up to the HII region edge which we define by the radius where the HII abundance is equal to 0.5. We then use the total C$^{+}$ mass fraction over the entire HII region, denoted as $f_{\rm{[CII]}}$, as a proxy for the expected reduction in [CII] luminosity for an HII region with the same physical conditions, if carbon was assumed to be entirely in the form of C$^{+}$. Figure \ref{fig: CLOUDY model} shows $f_{\rm{[CII]}}$ as a function of gas density for different radiation sources. For both single stars and star clusters, we find that $f_{\rm{[CII]}}$ decreases with increasing source luminosity, as well as with increasing gas density. For single stars, the hotter, more massive stars have a harder spectrum, resulting in a larger budget of C$^+$ ionizing photons, leading to a larger region where carbon is in the form of higher ionization stages. The density effect, as well as the luminosity effect in clusters, is a geometric effect, in which the mass of the thin C${^+}$ shell at the edge of the HII region increases (decreases) as the HII region size increases (decreases) with increasing luminosity (density). For an HII region density of 30 cm$^{-3}$, as is characteristic of our simulations, we find that $f_{\rm{[CII]}}$ ranges from 0.15 for a 45,000 K O-type star, to $\sim1$ for a 30,000 K B-type star. For a typical luminosity of an O-B association of $10^5\;L_{\odot}$, we find that $f_{\rm{[CII]}}=0.3$, and for all radiation sources this value decreases with increasing density. This simple model is in general agreement with the findings of \cite{Ebagezio2022}, and supports the claim that our [CII] emission associated with HII regions is a good approximation to within a factor of $\lesssim2$.}

\end{document}